\documentclass[preprint,prd,aps,showpacs,showkeys,nofootinbib]{revtex4}
\usepackage{amsmath}
\usepackage{graphicx}
\usepackage{caption}
\usepackage{subcaption}
\usepackage{makecell}
\usepackage{bm}
\usepackage{url}
\usepackage{float}
\UseRawInputEncoding

\begin{document}

\title{Probing the light charged Higgs boson, pseudoscalar Higgs boson, and $Z^\prime$ boson in the $U(1)_F$ model at the LHC}
\author{Zhan Cao$^{1,2,3}$\footnote{zhancao2022@163.com}, Zhong-Jun Yang$^{4}$\footnote{zj\_yang@lit.edu.cn}, Jin-Lei Yang$^{1,2,3}$\footnote{jlyang@hbu.edu.cn},  Tai-Fu Feng$^{1,2,3}$\footnote{fengtf@hbu.edu.cn}}

\affiliation{Department of Physics, Hebei University, Baoding, 071002, China$^1$\\
	Key Laboratory of High-precision Computation and Application of Quantum Field Theory of Hebei Province, Baoding, 071002, China$^2$\\
	Research Center for Computational Physics of Hebei Province, Baoding, 071002, China$^3$\\
    Department of mathematics and Physics, Luoyang institute of science and technology, Luoyang 471023, People's Republic of China$^4$}

\vspace*{-1cm}
\begin{abstract}
 In this papar, we study the production and decay of a charged Higgs boson, a pseudoscalar Higgs boson, and a $Z'$ boson at the LHC within the flavor-dependent model (FDM), at the LHC. Considering the constraints from perturbative unitarity and experimental measurements (e.g., the flavor physics data, higgs signal strengths, electroweak precision observables), we investigate the relevant processes by analyzing several common LHC search channels. Motivated by the excess in $t \to b\bar{b}c$ reported by ATLAS, which suggests a charged Higgs boson with a mass near 130 GeV and consistent with B-anomaly expectations, we perform a dedicated simulation for a charged Higgs around this mass. Our results support the experimental hint and predict that this particle has a high discovery potential at the future High-Luminosity Large Hadron Collider (HL-LHC). In contrast, for the pseudoscalar and $Z'$ bosons predicted in our model, they remain beyond the reach of current experiments as well as the expected sensitivity at a 14 TeV collider with an integrated luminosity of 300 fb$^{-1}$.
\end{abstract}

\keywords{FDM, Collider phenomenology}

\maketitle
\tableofcontents
\section{Introduction}\label{sec1}

The LHC, with its high energy, luminosity, and sensitivity, offers an effective experimental platform for probing phenomena in high-energy particle physics~\cite{Bruning:2025pmh,Bruning:2024zai}. For instance, the observation of the 125 GeV Higgs boson in 2012 provided a landmark confirmation of the Standard Model (SM)~\cite{ATLAS1,CMS1}. This observed Higgs boson state, however, could also potentially belong to the models with extended Higgs sectors. Such models, e.g., including extra doublets or triplets, remain viable possibilities that have not been ruled out~\cite{Arhrib:2020tqk}. The discovery of such new higgs bosons, if realized, would signal the existence of multiple Higgs sectors~\cite{Mondal:2023wib}. Some straightforward extension of the SM involves introducing an additional U(1) gauge group, and the corresponding gauge boson, commonly denoted as $Z^\prime$, is an electrically neutral spin-1 particle~\cite{Carena:2004xs,Feng:2016wph}. To satisfy stringent experimental bounds on the $Z^\prime$ mass, this U(1) symmetry must be spontaneously broken at a scale well above the electroweak scale~\cite{ParticleDataGroup:2024cfk,Basso:2010pe,Das:2021esm}. The discovery of any of these new particles would offer clearer evidence for physics beyond the SM~\cite{A:2025ygb,Accomando:2019ahs,Akeroyd:2016ymd}.

Since its proposal, SM has undergone extensive experimental validation and demonstrated remarkable success. Nevertheless, it has gradually revealed significant shortcomings. For instance, the quark flavor mixing described by the Cabibbo-Kobayashi-Maskawa (CKM) matrix~\cite{Cabibbo:1963yz,Kobayashi:1973fv} is not derived from first principles within the SM, which is known as the flavor puzzle. Moreover, the three generations of fermions exhibit strikingly different mass hierarchies, yet the SM offers no explanation for this pattern
\begin{eqnarray}
&&\frac{m_t}{m_c}\approx104,\;\frac{m_c}{m_u}\approx773,\;\frac{m_b}{m_s}\approx51,\;\frac{m_s}{m_d}\approx20,\;\frac{m_\tau}{m_\mu}\approx17,\;\frac{m_\mu}{m_e}\approx207.\label{eq1}
\end{eqnarray}
The mass hierarchy puzzle arises from the large differences in masses across the three families, despite sharing the same SM gauge group quantum numbers~\cite{ParticleDataGroup:2024cfk}. Furthermore, nonzero neutrino masses and mixings observed in oscillation experiments~\cite{Esteban:2018azc,Mirizzi:2015eza} suggest a mechanism beyond the SM for neutrino mass generation~\cite{Mohapatra:2005wg,Schechter:1980gr}. Additionally, there is the issue of dark matter, for which the SM does not provide a viable candidate particle~\cite{Arcadi:2017kky}. The FDM model, proposed as an extension of the SM, was designed to account for these issues~\cite{Yang:2024kfs}.

However, the observed consistency of the measured properties of the SM-like Higgs boson with predictions does not rule out the existence of additional scalars, such as a charged Higgs boson or a heavy neutral scalar Higgs boson. These could exist even at the electroweak scale in scenarios with small mixing angles\cite{CMS:2018cyk,Coloretti:2023yyq,LEPWorkingGroupforHiggsbosonsearches:2003ing,Crivellin:2021ubm}. Indeed, several experimental results have reported statistically significant hints (∼3$\sigma$) of a Higgs-like boson at the electroweak scale, including an excess in the $t \to b H^+$ process with $H^+ \to \bar{b}c$ and $M_{H^\pm} \approx 130\text{GeV}$~\cite{ATLAS:2023bzb,Coloretti:2025yji}, consistent with the expectations from the B-anomalies, i.e., $R_{D^{(\ast)}}$ and $b \to s\ell^+\ell^−$~\cite{HeavyFlavorAveragingGroupHFLAV:2024ctg,Hurth:2023jwr}. The FDM is precisely such a suitable model, as a non-minimal flavor structure is required to explain these observables~\cite{Coloretti:2025yji}. Additionally, as the FDM model introduces several new scalar particles, simulating their production and decay at the LHC provides an effective approach to investigate the extended Higgs sector~\cite{Li:2024kpd}, uncover potential signs of new physics (NP), and constrain the parameter space of the model. With the ongoing increases in luminosity and energy at the LHC, as well as the construction of future colliders, we can expect to continually deepen our understanding of these potentially existing scalar particles~\cite{Sopczak:2025ptc,ATLAS:2025cya}. 

The paper is organised as follows. In the next Section, we briefly describe the FDM model in Sec.~\ref{sec2}, then we describe the theoretical and experimental constraints to which the FDM is subjected in Sec.~\ref{sec3}. In the Sec.~\ref{sec4}, we introduce six Benchmark Points (BPs) for our Monte Carlo (MC) analysis which pass all present theoretical and experimental constraints. Finally, We discuss our analysis in Sec.~\ref{sec5} and we conclude in Sec.~\ref{sec6}.

\section{The flavor-dependent model}\label{sec2}

\begin{table*}
	\begin{tabular*}{\textwidth}{@{\extracolsep{\fill}}lllll@{}}
		\hline
		Multiplets & $SU(3)_C$ & $SU(2)_L$ & $U(1)_Y$ & $U(1)_F$\\
		\hline
		$l_1=(\nu_{1L},e_{1L})^T$ & 1 & 2 & $-\frac{1}{2}$ & $z$\\
		$l_2=(\nu_{2L},e_{2L})^T$ & 1 & 2 & $-\frac{1}{2}$ & $-z$\\
		$l_3=(\nu_{3L},e_{3L})^T$ & 1 & 2 & $-\frac{1}{2}$ & $0$\\
		$\nu_{1R}$                & 1 & 1 & $0$            & $-z$ \\
		$\nu_{2R}$                & 1 & 1 & $0$            & $z$ \\
		$e_{1R}$                  & 1 & 1 & $-1$           & $-z$ \\
		$e_{2R}$                  & 1 & 1 & $-1$           & $z$ \\
		$e_{3R}$                  & 1 & 1 & $-1$           & $0$ \\
		$q_1=(u_{1L},d_{1L})^T$   & 3 & 2 & $\frac{1}{6}$  & $z$\\
		$q_2=(u_{2L},d_{2L})^T$   & 3 & 2 & $\frac{1}{6}$  & $-z$\\
		$q_3=(u_{3L},d_{3L})^T$   & 3 & 2 & $\frac{1}{6}$  & $0$\\
		$d_{1R}$                  & 3 & 1 & -$\frac{1}{3}$ & $-z$ \\
		$d_{2R}$                  & 3 & 1 & -$\frac{1}{3}$ & $z$ \\
		$d_{3R}$                  & 3 & 1 & -$\frac{1}{3}$ & $0$ \\
		$u_{1R}$                  & 3 & 1 & $\frac{2}{3}$  & $-z$ \\
		$u_{2R}$                  & 3 & 1 & $\frac{2}{3}$  & $z$ \\
		$u_{3R}$                  & 3 & 1 & $\frac{2}{3}$  & $0$ \\
		$\Phi_1=(\phi_1^{+},\phi_1^{0})^T$   & 1 & 2 & $\frac{1}{2}$  & $z$\\
		$\Phi_2=(\phi_2^{+},\phi_2^{0})^T$   & 1 & 2 & $\frac{1}{2}$  & $-z$\\
		$\Phi_3=(\phi_3^{+},\phi_3^{0})^T$   & 1 & 2 & $\frac{1}{2}$  & 0\\
		$\chi$   & 1 & 1 & 0  & $2z$\\
		\hline
	\end{tabular*}
	\caption{Matter content in the FDM, where the nonzero constant $z$ denotes the extra $U(1)_F$ charge.}
	\label{tab2}
\end{table*}

The gauge group of the FDM is $SU(3)_C\otimes SU(2)_L\otimes U(1)_Y\otimes U(1)_F$, where the extra $U(1)_F$ local gauge group is related to the particles' flavor~\cite{Yang:2024znv}. Within this framework, fermion mass generation exhibits a hierarchical pattern: the third-generation fermions acquire their masses directly through tree-level Yukawa interactions, whereas the first and second generations obtain masses via tree-level mixing with the third generation. Realizing such intergenerational couplings requires the introduction of two additional scalar doublets~\cite{Yang:2024kfs}. To account simultaneously for the observed neutrino oscillations, the model further introduces two right-handed neutrinos and a scalar singlet. Upon spontaneous symmetry breaking, the scalar singlet develops a large vacuum expectation value, thereby inducing sizable Majorana masses for the right-handed neutrinos. Consequently, the light neutrino masses arise naturally through the conventional Type-I seesaw mechanism~\cite{Yang:2024znv}.

The field content of the FDM and their corresponding gauge charges are summarized in Table~\ref{tab2}, where the nonzero parameter \( z \) represents the additional \( U(1)_F \) charge. Notably, the model contains only two generations of right-handed neutrinos: since both the \( U(1)_F \) and \( U(1)_Y \) charges of the third-generation right-handed neutrino \( \nu_{R_3} \) vanish, it becomes physically redundant and is omitted. Furthermore, the fermionic charge assignments in Table~\ref{tab2} ensure the cancellation of all chiral anomalies.

\subsection{The scalar sector of the FDM}\label{sec2-1}

The scalar potential in the FDM can be written as~\cite{Yang:2024znv,Yang:2024kfs}
\begin{eqnarray}
	&&V=-M_{\Phi_1}^2 \Phi_1^\dagger\Phi_1-M_{\Phi_2}^2 \Phi_2^\dagger\Phi_2-M_{\Phi_3}^2 \Phi_3^\dagger\Phi_3-M_{\chi}^2\chi^*\chi+\lambda_{\chi} (\chi^*\chi)^2+\lambda_1 (\Phi_1^\dagger\Phi_1)^2\nonumber\\
	&&\qquad+\lambda_2 (\Phi_2^\dagger\Phi_2)^2+\lambda_3 (\Phi_3^\dagger\Phi_3)^2+\lambda'_4 (\Phi_1^\dagger\Phi_1)(\Phi_2^\dagger\Phi_2)+\lambda_4'' (\Phi_1^\dagger\Phi_2)(\Phi_2^\dagger\Phi_1)\nonumber\\
	&&\qquad+\lambda_5' (\Phi_1^\dagger\Phi_1)(\Phi_3^\dagger\Phi_3)+\lambda_5'' (\Phi_1^\dagger\Phi_3)(\Phi_3^\dagger\Phi_1)+\lambda_6' (\Phi_2^\dagger\Phi_2)(\Phi_3^\dagger\Phi_3)+\lambda_6'' (\Phi_2^\dagger\Phi_3)(\Phi_3^\dagger\Phi_2)\nonumber\\
	&&\qquad+\lambda_7 (\Phi_1^\dagger\Phi_1)(\chi^*\chi)+\lambda_{8} (\Phi_2^\dagger\Phi_2)(\chi^*\chi)+\lambda_{9} (\Phi_3^\dagger\Phi_3)(\chi^*\chi)+[\lambda_{10} (\Phi_3^\dagger\Phi_1)(\Phi_3^\dagger\Phi_2)\nonumber\\
	&&\qquad+\kappa(\Phi_1^\dagger\Phi_2)\chi+h.c.],\label{eqsca}
\end{eqnarray}
where
\begin{align}
	\Phi_1&=\left(\begin{array}{c}\phi_1^+\\ \frac{1}{\sqrt2}(i A_1+S_1+v_1)\end{array}\right), &
	\Phi_2&=\left(\begin{array}{c}\phi_2^+\\ \frac{1}{\sqrt2}(i A_2+S_2+v_2)\end{array}\right), \nonumber\\
	\Phi_3&=\left(\begin{array}{c}\phi_3^+\\ \frac{1}{\sqrt2}(i A_3+S_3+v_3)\end{array}\right), &
	\chi&=\frac{1}{\sqrt2}(i A_{\chi}+S_{\chi}+v_\chi),
\end{align}
and $v_i\;(i=1,\;2,\;3),\;v_\chi$ are the VEVs of $\Phi_i,\;\chi$ respectively.

Based on the scalar potential in Eq.~(\ref{eqsca}), the tadpole equations in the FDM can be expressed as
\begin{eqnarray}
	&&M_{\Phi_1}^2=\lambda_1v_1^2+\frac{1}{2}\Big[(\lambda_4'+\lambda_4'') v_2^2+(\lambda_5'+\lambda_5'') v_3^2+\frac{v_2}{v_1}v_3^2 {\rm Re}(\lambda_{10})+\sqrt2\frac{v_2}{v_1} v_\chi {\rm Re}(\kappa)+\lambda_7v_\chi^2\Big],\nonumber\\
	&&M_{\Phi_2}^2=\lambda_2v_2^2+\frac{1}{2}\Big[(\lambda_4'+\lambda_4'') v_1^2+(\lambda_6'+\lambda_6'') v_3^2+\frac{v_1}{v_2}v_3^2 {\rm Re}(\lambda_{10})+\sqrt2\frac{v_1}{v_2} v_\chi {\rm Re}(\kappa)+\lambda_8v_\chi^2\Big],\nonumber\\
	&&M_{\Phi_3}^2=\lambda_3v_3^2+{\rm Re}(\lambda_{10})v_1v_2+\frac{1}{2}[(\lambda_5'+\lambda_5'') v_1^2+(\lambda_6'+\lambda_6'') v_2^2+\lambda_9 v_c^2],\nonumber\\
	&&M_{\chi}^2=\lambda_\chi v_\chi^2+\frac{1}{2}\Big[\lambda_7 v_1^2+\lambda_8 v_2^2+\lambda_9 v_3^2+\sqrt2\frac{v_1v_2}{v_\chi}  {\rm Re}(\kappa)\Big].\label{eqtad}
\end{eqnarray}

On the basis $(S_1,\;S_2,\;S_3,\;S_\chi)$, the CP-even Higgs squared mass matrix in the FDM is
\begin{eqnarray}
	&&M_{h}^2=\left(\begin{array}{*{20}{cccc}}
		M_{h,11}^2 & M_{h,12}^2 & M_{h,13}^2 & M_{h,14}^2 \\ [6pt]
		M_{h,12}^2 & M_{h,22}^2 & M_{h,23}^2 & M_{h,24}^2 \\ [6pt]
		M_{h,13}^2 & M_{h,23}^2 & M_{h,33}^2 & M_{h,34}^2 \\ [6pt]
		M_{h,14}^2 & M_{h,24}^2 & M_{h,34}^2 & M_{h,44}^2 \\ [6pt]
	\end{array}\right),\label{eq4h}
\end{eqnarray}
where
\begin{eqnarray}
	&&M_{h,11}^2=2\lambda_1v_1^2-\frac{v_2}{2v_1}\Big[v_3^2 {\rm Re}(\lambda_{10})+\sqrt2 v_\chi {\rm Re}(\kappa)\Big],\nonumber\\
	&&M_{h,22}^2=2\lambda_2v_2^2-\frac{v_1}{2v_2}\Big[v_3^2 {\rm Re}(\lambda_{10})+\sqrt2 v_\chi {\rm Re}(\kappa)\Big],\nonumber\\
	&&M_{h,33}^2=2\lambda_3v_3^2,\;\;M_{h,44}^2=2\lambda_\chi v_\chi^2-\frac{\sqrt2v_1v_2}{2v_\chi}  {\rm Re}(\kappa),\nonumber\\
	&&M_{h,12}^2=(\lambda_4'+\lambda_4'') v_1v_2+\frac{1}{2}{\rm Re}(\lambda_{10})v_3^2+\frac{\sqrt2}{2}{\rm Re}(\kappa)v_\chi,\nonumber\\
	&&M_{h,13}^2=(\lambda_5'+\lambda_5'')v_1v_3+{\rm Re}(\lambda_{10})v_2v_3,\;\;M_{h,14}^2=\lambda_7v_1v_\chi+\frac{\sqrt2}{2}{\rm Re}(\kappa)v_2,\nonumber\\
	&&M_{h,23}^2=(\lambda_6'+\lambda_6'')v_2v_3+{\rm Re}(\lambda_{10})v_1v_3,\;\;M_{h,24}^2=\lambda_8v_2v_\chi+\frac{\sqrt2}{2}{\rm Re}(\kappa)v_1,\nonumber\\
	&&M_{h,34}^2=\lambda_9v_3v_\chi.\label{eqmh}
\end{eqnarray}

The tadpole equations in Eq.~(\ref{eqtad}) are used to obtain the matrix elements above.

Then, on the basis $(A_1,\;A_2,\;A_3,\;A_\chi)$, the squared mass matrix of CP-odd Higgs in the FDM can be written as
\begin{eqnarray}
&&M_{A}^2=\left(\begin{array}{*{20}{cccc}}
M_{A,11}^2 & M_{A,12}^2 & M_{A,13}^2 & M_{A,14}^2 \\ [6pt]
M_{A,12}^2 & M_{A,22}^2 & M_{A,23}^2 & M_{A,24}^2 \\ [6pt]
M_{A,13}^2 & M_{A,23}^2 & M_{A,33}^2 & M_{A,34}^2 \\ [6pt]
M_{A,14}^2 & M_{A,24}^2 & M_{A,34}^2 & M_{A,44}^2 \\ [6pt]
\end{array}\right),
\end{eqnarray}
where
\begin{eqnarray}
&&M_{A,11}^2=-\frac{v_2}{2v_1}[{\rm Re}(\lambda_{10})v_3^2+\sqrt2 v_\chi {\rm Re} (\kappa)],\;\;M_{A,33}^2=-2{\rm Re}(\lambda_{10})v_1v_2,\nonumber\\
&&M_{A,22}^2=-\frac{v_1}{2v_2}[{\rm Re}(\lambda_{10})v_3^2+\sqrt2 v_\chi {\rm Re} (\kappa)],\;\;M_{A,44}^2=-\frac{\sqrt2 v_1v_2}{2v_\chi}{\rm Re} (\kappa),\nonumber\\
&&M_{A,12}^2=\frac{\sqrt 2}{2} v_\chi {\rm Re} (\kappa)-\frac{1}{2}{\rm Re}(\lambda_{10})v_3^2,\;\;M_{A,13}^2={\rm Re}(\lambda_{10})v_2v_3,\nonumber\\
&&M_{A,14}^2=\frac{\sqrt 2}{2} v_2 {\rm Re} (\kappa),\;\;M_{A,23}^2={\rm Re}(\lambda_{10})v_1v_3,\;\;M_{A,24}^2=-\frac{\sqrt 2}{2} v_1 {\rm Re} (\kappa),\nonumber\\
&&M_{A,34}^2=0.\label{eqmA}
\end{eqnarray}

On the basis $(\phi_1^+,\;\phi_2^+,\;\phi_3^+)$ and $(\phi_1^-,\;\phi_2^-,\;\phi_3^-)^T$, the squared mass matrix of singly charged Higgs in the FDM can be written as
\begin{eqnarray}
	&&M_{H^\pm}^2=\left(\begin{array}{*{20}{ccc}}
		M_{H^\pm,11}^2 & M_{H^\pm,12}^2 & M_{H^\pm,13}^2 \\ [6pt]
		(M_{H^\pm,12}^2)^* & M_{H^\pm,22}^2 & M_{H^\pm,23}^2 \\ [6pt]
		(M_{H^\pm,13}^2)^* & (M_{H^\pm,23}^2)^* & M_{H^\pm,33}^2 \\ [6pt]
	\end{array}\right),
\end{eqnarray}
where
\begin{eqnarray}
	&&M_{H^\pm,11}^2=-\frac{v_2}{2v_1}[{\rm Re}(\lambda_{10})v_3^2+\sqrt2 v_\chi {\rm Re} (\kappa)]-\frac{1}{2}(\lambda_4'' v_2^2+\lambda_5'' v_3^2),\nonumber\\
	&&M_{H^\pm,22}^2=-\frac{v_1}{2v_2}[{\rm Re}(\lambda_{10})v_3^2+\sqrt2 v_\chi {\rm Re} (\kappa)]-\frac{1}{2}(\lambda_4'' v_1^2+\lambda_6'' v_3^2),\nonumber\\
	&&M_{H^\pm,33}^2=-{\rm Re}(\lambda_{10})v_1v_2-\frac{1}{2}(\lambda_5'' v_1^2+\lambda_6'' v_2^2),\nonumber\\
	&&M_{H^\pm,12}^2=\frac{\sqrt 2}{2}v_\chi\kappa+\frac{1}{2}\lambda_4'' v_1 v_2,\;\;M_{H^\pm,13}^2=\frac{1}{2}v_3(\lambda_5'' v_1+\lambda_{10}^*v_2),\nonumber\\
	&&M_{H^\pm,23}^2=\frac{1}{2}v_3(\lambda_6'' v_2+\lambda_{10}^*v_1).\label{eqmCH}
\end{eqnarray}
It is easy to verify that there are two neutral Goldstones and one singly charged Goldstone in the FDM.

\subsection{The fermion masses in the FDM}\label{sec2-2}

Based on the matter content listed in Tab.~\ref{tab2}, the Yukawa couplings in the FDM can be written as
\begin{eqnarray}
	&&\mathcal{L}_Y=Y_u^{33}\bar q_3 \tilde \Phi_3 u_{R_3}+Y_d^{33}\bar q_3 \Phi_3 d_{R_3}+Y_u^{32}\bar q_3 \tilde{\Phi}_1 u_{R_2}+Y_u^{23}\bar q_2 \tilde \Phi_1 u_{R_3}+Y_d^{32}\bar q_3 \Phi_2 d_{R_2}\nonumber\\
	&&\qquad\; +Y_d^{23}\bar q_2 \Phi_2 d_{R_3}+Y_u^{21}\bar q_2 \tilde{\Phi}_3 u_{R_1}+Y_u^{12}\bar q_1 \tilde \Phi_3 u_{R_2}+Y_d^{21}\bar q_2 \Phi_3 d_{R_1}+ Y_d^{12}\bar q_1 \Phi_3 d_{R_2}\nonumber\\
	&&\qquad\; +Y_u^{31}\bar q_3 \tilde{\Phi}_2 u_{R_1}+Y_u^{13}\bar q_1 \tilde \Phi_2 u_{R_3}+Y_d^{31}\bar q_3 \Phi_1 d_{R_1}+Y_d^{13}\bar q_1 \Phi_1 d_{R_3}\nonumber\\
	&&\qquad\; +Y_e^{33}\bar l_3 \Phi_3 e_{R_3}+Y_e^{32}\bar l_3 \Phi_2 e_{R_2}+Y_e^{23}\bar l_2 \Phi_2 e_{R_3}+Y_e^{21}\bar l_2 \Phi_3 e_{R_1}+ Y_e^{12}\bar l_1 \Phi_3 e_{R_2}\nonumber\\
	&&\qquad\; +Y_e^{31}\bar l_3 \Phi_1 e_{R_1}+Y_e^{13}\bar l_1 \Phi_1 e_{R_3}+Y_R^{11}\bar\nu^c_{R_1}\nu_{R_1}\chi+Y_R^{22}\bar\nu^c_{R_2}\nu_{R_2} \chi^*+Y_D^{21}\bar l_2 \tilde \Phi_3 \nu_{R_1}\nonumber\\
	&&\qquad\; +Y_D^{12}\bar l_1 \tilde \Phi_3 \nu_{R_2}+Y_D^{31}\bar l_3 \tilde \Phi_2 \nu_{R_1}+Y_D^{32}\bar l_3 \tilde \Phi_1 \nu_{R_2}+h.c..\label{eq9}
\end{eqnarray}
Then the mass matrices of quarks and leptons can be written as~\cite{Yang:2024znv,Yang:2024kfs}
\begin{eqnarray}
	&&m_q=\left(\begin{array}{ccc} 0 & m_{q,12} & m_{q,13}\\
		m_{q,12}^* & 0 & m_{q,23}\\
		m_{q,13}^* & m_{q,23}^* & m_{q,33}\end{array}\right),m_e=\left(\begin{array}{ccc} 0 & m_{e,12} & m_{e,13}\\
		m_{e,12}^* & 0 & m_{e,23}\\
		m_{e,13}^* & m_{e,23}^* & m_{e,33}\end{array}\right),m_\nu=\left(\begin{array}{cc} 0 & M_D^T\\
		M_D & M_R\end{array}\right).\label{eq2}
\end{eqnarray}
Where $q=u,d$, the parameters $m_{q,33}$ and $m_{e,33}$ are real, $M_D$ is $2\times3$ Dirac mass matrix and $M_R$ is $2\times2$ Majorana mass matrix (the nonzero neutrino masses are obtained by the Type I see-saw mechanism). The elements of the matrices in Eq.~(\ref{eq2}) are
\begin{eqnarray}
	&&m_{u,11}=m_{u,22}=0,\;m_{u,33}=\frac{1}{\sqrt2}Y_u^{33}v_3,\;m_{u,12}=\frac{1}{\sqrt2}Y_u^{12}v_3,\;m_{u,13}=\frac{1}{\sqrt2}Y_u^{13}v_1,\nonumber\\
	&&m_{u,23}=\frac{1}{\sqrt2}Y_u^{23}v_2,\label{eqmu}\\
	&&m_{d,11}=m_{d,22}=0,\;m_{d,33}=\frac{1}{\sqrt2}Y_d^{33}v_3,\;m_{d,12}=\frac{1}{\sqrt2}Y_d^{12}v_3,\;m_{d,13}=\frac{1}{\sqrt2}Y_d^{13}v_1,\nonumber\\
	&&m_{d,23}=\frac{1}{\sqrt2}Y_d^{23}v_2,\label{eqmd}\\
	&&m_{e,11}=m_{e,22}=0,\;m_{e,33}=\frac{1}{\sqrt2}Y_e^{33}v_3,\;m_{e,12}=\frac{1}{\sqrt2}Y_e^{12}v_3,\;m_{e,13}=\frac{1}{\sqrt2}Y_e^{13}v_1,\nonumber\\
	&&m_{e,23}=\frac{1}{\sqrt2}Y_e^{23}v_2,\label{eqme}\\
	&&M_{D,11}=M_{D,22}=0,\;\;M_{D,12}=\frac{1}{\sqrt2}Y_D^{12}v_3,\;M_{D,31}=\frac{1}{\sqrt2}Y_D^{31}v_1,\nonumber\\
	&&M_{D,32}=\frac{1}{\sqrt2}Y_D^{32}v_2,\;M_{R,12}=M_{R,21}=0,\;M_{R,11}=\frac{1}{\sqrt2}Y_R^{11}v_\chi,\;M_{R,22}=\frac{1}{\sqrt2}Y_R^{22}v_\chi.
\end{eqnarray}

\subsection{The gauge sector of the FDM}\label{sec2-3}

Due to the introducing of an extra $U(1)_F$ local gauge group in the FDM, the covariant derivative corresponding to $SU(2)_L\otimes U(1)_Y\otimes U(1)_F$ is defined as~\cite{Yang:2024znv,Yang:2024kfs}
\begin{eqnarray}
	&&D_\mu=\partial_\mu+i g_2 T_j A_{j\mu}+i g_1 Y B_\mu+i g_{_F} F B'_\mu+i g_{_{YF}} Y B'_\mu,\;(j=1,\;2,\;3),\label{eqCD}
\end{eqnarray}
where $(g_2,\;g_1,\; g_{_F})$, $(T_j,\;Y,\;F)$, $(A_{j\mu},\;B_\mu,\; B'_\mu)$ denote the gauge coupling constants, generators and gauge bosons of groups $(SU(2)_L,\;U(1)_Y,\;U(1)_F)$ respectively. The coupling $g_{YF}$ arises from the gauge kinetic mixing effect which presents in the models with two Abelian groups. Then the W boson mass can be written as
\begin{eqnarray}
	&&M_W=\frac{1}{2}g_2 (v_1^2+v_2^2+v_3^2)^{1/2},
\end{eqnarray}
where $(v_1^2+v_2^2+v_3^2)^{1/2}=v\approx246.22\;{\rm GeV}$ and we have $v_1=\;v_2 < v_3$ in the FDM. The $\gamma$, $Z$ and $Z^\prime$ boson masses in the FDM can be written as
\begin{eqnarray}
	&&M_\gamma=0,\;M_Z\approx\frac{1}{2}(g_1^2+g_2^2)^{1/2} v,\;M_{Z'}\approx 2|zg_{_F}| v_\chi,\label{eq19}
\end{eqnarray}
and
\begin{eqnarray}
	&&\gamma=c_W B+s_W A_3,\;Z=-s_W B+c_W A_3+s'_W B',\;Z'=s_W'(s_WB-c_W A_3)+c'_W B',
\end{eqnarray}
where $\gamma,\;Z,\;Z'$ are the mass eigenstates, $c_W\equiv \cos \theta_W,\;s_W\equiv \sin \theta_W$ with $\theta_W$ denoting the Weinberg angle, $s_W'\equiv \sin \theta'_W$, $c_W'\equiv \cos \theta'_W$ with $\theta_W'$ representing the $Z-Z'$ mixing effect.

\section{Bounds and constraints in the parameter space}\label{sec3}

There are many free parameters in the FDM. To obtain tractable numerical results while maintaining generality, we adopt the following simplifying assumptions for the scalar sector parameters: $v_1=v_2=v_{1,2}$, $\lambda_4'=\lambda_4''=\lambda_4/2$, $\lambda_5'=\lambda_5''=\lambda_5/2$, $\lambda_6'=\lambda_6''=\lambda_6/2$. The Higgs sector properties show negligible dependence on the gauge couplings $g_F$, $g_{YF}$ hence we take $g_F=0.4$, $g_{YF}=0.2$ in the computations. Furthermore, the vacuum expectation value $v_\chi$ is constrained by the mass of the new $Z^\prime$ boson through Eq.~(\ref{eq19}), leading us to adopt $v_\chi\geq4\;{\rm TeV}$ in the subsequent analysis.

\subsection{The electroweak precision observables}\label{sec3-1}

We have taken into account the constraints from electroweak precision observables through the oblique parameters S, T, and U, and obtained the following results~\cite{ParticleDataGroup:2024cfk},
\begin{equation}
\mathrm{S}=-0.04\pm0.10,\mathrm{T}=0.01\pm0.12,\mathrm{U}=-0.01\pm0.11.
\end{equation}

Then, we use SPheno~\cite{Porod:2003um} to compute the mass spectrum and these parameters, and subsequently pass the results to Higgstools-1.1.3 program~\cite{Bahl:2022igd}.

\subsection{Vacuum stability, Perturbativity constraints, Signal strengths}\label{sec3-2}

To accommodate potential additional Higgs bosons while maintaining consistency with the measurements of the observed Higgs state, whose properties align with the SM predictions—we impose constraints using the Higgstools. This tool incorporates combined measurements of the SM-like Higgs boson from LHC Run-1 and Run-2 data, and systematically evaluates each parameter point against 95\% confidence level (CL) exclusion limits. These limits are derived from Higgs boson searches performed at LEP, the Tevatron, and the LHC experiments. The vacuum stability requirement the scalar potential should be bounded by the conditions
\begin{eqnarray}
&&\lambda_1> 0,\lambda_2> 0,\lambda_3> 0,\lambda_\chi > 0,\left | \lambda_{10} \right |^2\le (L_{13} +2 \sqrt{\lambda_1 \lambda_3})(L_{23} +2 \sqrt{\lambda_2 \lambda_3}),\nonumber\\
&&L_{12} +2 \sqrt{\lambda_1 \lambda_2}\ge 0,L_{13} +2 \sqrt{\lambda_1 \lambda_3}\ge 0,L_{23} +2 \sqrt{\lambda_2 \lambda_3}\ge 0,\nonumber\\
&&\lambda_7 +2 \sqrt{\lambda_1 \lambda_\chi}\ge 0,\lambda_8 +2 \sqrt{\lambda_2 \lambda_\chi}\ge 0,\lambda_9 +2 \sqrt{\lambda_3 \lambda_\chi}\ge 0,\nonumber\\
&&\sqrt{\lambda_1 \lambda_2 \lambda_\chi} +\frac{L_{12} \sqrt{\lambda_\chi}+\lambda_7 \sqrt{\lambda_2}+\lambda_8 \sqrt{\lambda_1}}{2}+\frac{1}{2} \sqrt{(L_{12}+2\sqrt{\lambda_1 \lambda_2})(\lambda_7+2\sqrt{\lambda_1 \lambda_\chi})(\lambda_8+2\sqrt{\lambda_2 \lambda_\chi})}\ge 0,\nonumber\\
&&\sqrt{\lambda_1 \lambda_3 \lambda_\chi} +\frac{L_{13} \sqrt{\lambda_\chi}+\lambda_7 \sqrt{\lambda_3}+\lambda_9 \sqrt{\lambda_1}}{2}+\frac{1}{2} \sqrt{(L_{13}+2\sqrt{\lambda_1 \lambda_3})(\lambda_7+2\sqrt{\lambda_1 \lambda_\chi})(\lambda_9+2\sqrt{\lambda_3 \lambda_\chi})}\ge 0,\nonumber\\
&&\sqrt{\lambda_2 \lambda_3 \lambda_\chi} +\frac{L_{23} \sqrt{\lambda_\chi}+\lambda_8\sqrt{\lambda_3}+\lambda_9 \sqrt{\lambda_2}}{2}+\frac{1}{2} \sqrt{(L_{23}+2\sqrt{\lambda_2 \lambda_3})(\lambda_8+2\sqrt{\lambda_2 \lambda_\chi})(\lambda_9+2\sqrt{\lambda_3 \lambda_\chi})} \ge 0,\nonumber\\
&&\sqrt{\lambda_1 \lambda_2 \lambda_3} +\frac{L_{12} \sqrt{\lambda_3}+\tilde{L}_{13}\sqrt{\lambda_2}+\tilde{L}_{23}\sqrt{\lambda_1}}{2}+\frac{1}{2} \sqrt{(L_{12}+2\sqrt{\lambda_1 \lambda_2})(\tilde{L}_{13}+2\sqrt{\lambda_1 \lambda_3})(\tilde{L}_{23}+2\sqrt{\lambda_2 \lambda_3})} \ge 0.\nonumber\\
\label{eqVS}
\end{eqnarray}
Where
\begin{eqnarray}
&&L_{12}=\lambda_4'+\text{min}(0,\lambda_4''),L_{13}=\lambda_5'+\text{min}(0,\lambda_5''),L_{23}=\lambda_6'+\text{min}(0,\lambda_6''),\nonumber\\
&&\tilde{L}_{13}=L_{13}-\left | \lambda_{10} \right | \sqrt{\frac{L_{13} +2 \sqrt{\lambda_1 \lambda_3}}{L_{23} +2 \sqrt{\lambda_2 \lambda_3}} },\nonumber\\
&&\tilde{L}_{23}=L_{23}-\left | \lambda_{10} \right | \sqrt{\frac{L_{23} +2 \sqrt{\lambda_2 \lambda_3}}{L_{13} +2 \sqrt{\lambda_1 \lambda_3}} }.
\label{eqLC}
\end{eqnarray}
We have the tree-level perturbative unitary constraints
\begin{eqnarray}
&&|\lambda_1|,\;|\lambda_2|,\;|\lambda_3|,\;|\lambda_\chi|\leq\frac{4\pi}{3},\;|{\rm Re}(\lambda_{10})|\leq4\pi,\nonumber\\
&&|\lambda_4'+\lambda_4''|,\;|\lambda_5'+\lambda_5''|,\;|\lambda_6'+\lambda_6''|,\;|\lambda_7|,\;|\lambda_8|,\;|\lambda_9|\leq 8\pi.\label{eqPU}
\end{eqnarray}

Simultaneously, we also included the latest experimental constraints on the signal strengths, which are listed in Table~\ref{tabsg}.

\begin{table*}
	\begin{tabular*}{\textwidth}{@{\extracolsep{\fill}}lllll@{}}
		\hline
		Signal & Value from PDG ~\cite{ParticleDataGroup:2024cfk}\\
		\hline
		$\mu_{\gamma\gamma}$ & $1.10\pm0.06$ \\
		$\mu_{ZZ^*}$ & $1.02\pm0.08$ \\
		$\mu_{WW^*}$ & $1.00\pm0.08$  \\
		$\mu_{b\bar{b}}$ & $0.99\pm0.12$ \\
		$\mu_{c\bar{c}}$ & $<14$ \\
		$\mu_{\tau\bar{\tau}}$ & $0.91\pm0.09$  \\
		$\mu_{Z\gamma}$ & $2.2\pm0.7$  \\
		\hline
	\end{tabular*}
	\caption{Experimental values for the Higgs decay rates.}
	\label{tabsg}
\end{table*}

\subsection{Constraints of flavor physics}\label{sec3-3}
The newly defined Z boson, particularly the $Z^\prime$ boson, can mediate Flavor-Changing Neutral Currents (FCNCs). Therefore, we consider some effects of FCNCs in FDM. We primarily consider the following processes: The $B$ meson rare decay processes $\bar B \to X_s\gamma$, $B_s^0 \to \mu^+\mu^-$ are related closely to the NP contributions, and the average experimental constraints on the branching ratios of $\bar B \to X_s\gamma$, $B_s^0 \to \mu^+\mu^-$ are
\begin{eqnarray}
&&{\rm Br}(\bar B \to X_s\gamma)=(3.49\pm0.19)\times 10^{-4},\nonumber\\
&&{\rm Br}(B_s^0 \to \mu^+\mu^-)=(3.01\pm0.35)\times 10^{-9}.\label{eqBD}
\end{eqnarray}

The newly introduced scalars in the FDM including CP-even Higgs bosons, CP-odd Higgs bosons and charged Higgs bosons can make contributions to these two processes, the analytical calculations of the contributions are collected in the Appendix of Ref.~\cite{Yang:2018fvw}

The branching ratios of the top quark rare decay processes $t\to ch$ and $t\to uh$ can be written as~\cite{Yang:2018utw}
\begin{eqnarray}
&&{\rm Br}(t\rightarrow q_u h)=\frac{|\mathcal{M}_{t q_u h}|^2\sqrt{((m_t+m_h)^2-m_{q_u}^2)((m_t-m_h)^2-m_{q_u}^2)}}{32\pi m_t^3\Gamma^t_{{\rm total}}},
\end{eqnarray}
where $q_u=u,\;c$, the amplitude $\mathcal{M}_{tq_uh}$ can be read directly from the Yukawa couplings in Eq.~(\ref{eq9}), and $\Gamma^t_{{\rm total}}=1.42\;$GeV~\cite{ParticleDataGroup:2024cfk} is the total decay width of top quark. 
The experimental constraints on the branching ratios of $t\to ch$, $t\to uh$ are
\begin{eqnarray}
&&{\rm Br}(t\to ch)<3.4\times 10^{-4},\nonumber\\
&&{\rm Br}(t\to uh)<1.9\times 10^{-9}.\label{eqBD1}
\end{eqnarray}

Finally, we focus on the lepton flavor violation processes $\tau\to 3e$, $\tau\to 3\mu$, $\mu\to 3e$ predicted in the FDM. The corresponding amplitude can be written as~\cite{Hisano:1995cp}
\begin{eqnarray}
&&\mathcal{M}(e_j\rightarrow e_i e_i\bar e_i)=C_1^L\bar u_{e_i}(p_2)\gamma_\mu P_L u_{e_j}(p_1) u_{e_i}(p_3)\gamma^\mu P_L \nu_{e_i}(p_4)\nonumber\\
&&\qquad\quad+C_1^R\bar u_{e_i}(p_2)\gamma_\mu P_R u_{e_j}(p_1) u_{e_i}(p_3)\gamma^\mu P_R \nu_{e_i}(p_4)\nonumber\\
&&\qquad\quad+[C_2^L\bar u_{e_i}(p_2)\gamma_\mu P_L u_{e_j}(p_1) u_{e_i}(p_3)\gamma^\mu P_R \nu_{e_i}(p_4)\nonumber\\
&&\qquad\quad+C_2^R\bar u_{e_i}(p_2)\gamma_\mu P_R u_{e_j}(p_1) u_{e_i}(p_3)\gamma^\mu P_L \nu_{e_i}(p_4)-(p_2\leftrightarrow p_3)]\nonumber\\
&&\qquad\quad+[C_3^L\bar u_{e_i}(p_2) P_L u_{e_j}(p_1) u_{e_i}(p_3) P_L \nu_{e_i}(p_4)\nonumber\\
&&\qquad\quad+C_3^R\bar u_{e_i}(p_2) P_R u_{e_j}(p_1) u_{e_i}(p_3) P_R \nu_{e_i}(p_4)-(p_2\leftrightarrow p_3)],
\end{eqnarray}
where $i=1,\;2$ for $j=3$, $i=1$ for $j=2$, $u_{e_i}$ denotes the spinor of lepton, $\nu_{e_i}$ denotes the spinor of antilepton, $P_L=(1-\gamma_5)/2$, $P_R=(1+\gamma_5)/2$, and $p_k$ denotes the momentum of charged lepton with $k=1,2,3,4$. The coefficients $C_{1,2,3}^{L,R}$ from the contributions of Higgs bosons and $Z,\;Z'$ gauge bosons, can be obtained through the Yukawa couplings in Eq.~(\ref{eq9}) and the definition of covariant derivative in Eq.~(\ref{eqCD}). Then we can calculate the decay rate~\cite{Hisano:1995cp}
\begin{eqnarray}
&&\Gamma(e_j\rightarrow e_i e_i\bar e_i)=\frac{m_{e_j}^5}{1536\pi^3}\Big[\frac{1}{2}(|C_1^L|^2+|C_1^R|^2)+|C_2^L|^2+|C_2^R|^2+\frac{1}{8}(|C_3^L|^2+|C_3^R|^2)\Big].
\end{eqnarray}
The total decay widthes of $\mu,\;\tau$ are taken as $\Gamma^\mu_{{\rm total}}=2.996\times 10^{-19}\;$GeV, $\Gamma^\tau_{{\rm total}}=2.265\times 10^{-12}\;$GeV~\cite{ParticleDataGroup:2024cfk}.

\begin{table}
	\begin{tabular*}{\textwidth}{@{\extracolsep{\fill}}lllll@{}}
		\hline
		Parameters&Min&Max\\
		\hline
		$v_1/{\rm GeV}$&30&142&\\
		$\lambda_i$ $\;(i = 1, 2, \cdots, 9,\chi)$&0&5\\
		$\lambda_{10}$&-4&0\\
        $v_{\chi}/{\rm GeV}$&4000&8000\\
        $\kappa/{\rm GeV}$&-3000&-100\\
		\hline
	\end{tabular*}
	\caption{Scanning the parameters of the scalar potential coefficients in the Higgs decays, imposing constraints from both vacuum stability and perturbative unitarity conditions.}
	\label{tab-scan}
\end{table}

After taking into account the constraints from vacuum stability and perturbative unitarity conditions, we select the initial parameter space as shown in Table~\ref{tab-scan} and keep the lightest neutral Higgs mass in the range of $124\;{\rm GeV}< m_h< 126\;{\rm GeV}$.

\section{Collider phenomenology}\label{sec4}
In this section, we present a detailed MC analysis for both signal and background events at the detector level. To test the allowed parts of the parameter space, six BPs are given in the Table ~\ref{t:BPs}.

\begin{table}[!ht]
	\begin{center}
			\begin{tabular}{| c| c| c| c| c| c| c|}
				\hline
				~ & $M_{Zp}$ & $M_A$ & $M_H^{\pm}$ & $v_3$ & $v_1$\\
				\hline  
BP1 & 6013.74 & 320.33 & 144.89 & 201.48 & 100.08 \\
\hline
BP2 & 5514.78 & 332.60 & 158.11 & 209.90 & 91.01 \\
\hline
BP3 & 5830.72 & 321.10 & 145.74 & 218.78 & 79.88 \\
\hline
BP4 & 6720.11 & 280.05 & 94.35 & 230.98 & 60.30 \\
\hline
BP5 & 6234.79 & 311.83 & 135.30 & 191.20 & 109.69\\
\hline
BP6 & 8012.29 & 293.67 & 113.17 & 174.20 & 123.04 \\
   				\hline
			\end{tabular}
			\caption{FDM input parameters are presented  for each BP. The unit of all masses is GeV and we fix the SM-like Higgs boson mass in the range of $124\;{\rm GeV}< m_h< 126\;{\rm GeV}$.}\label{t:BPs}
	\end{center}
\end{table}

\subsection{Charged Higgs Collider phenomenology}\label{sec4-1}
\begin{figure}[!ht]
  \centering
  \begin{subfigure}[t]{.28\textwidth}
    \centering
    \includegraphics[width=\linewidth]{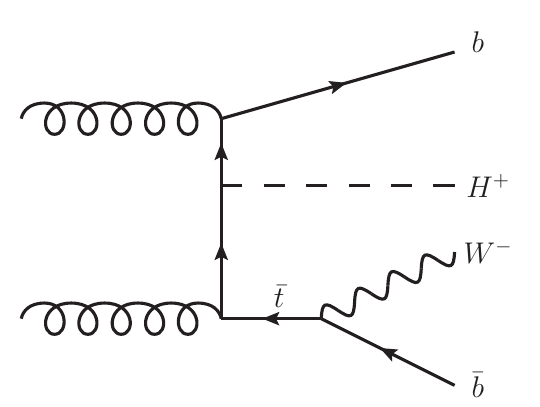}
    \caption{single-resonant top-quark production}
    \label{fig:placeholder}
  \end{subfigure}
  \hspace{0.04\textwidth}
  \begin{subfigure}[t]{.28\textwidth}
    \centering
    \includegraphics[width=\linewidth]{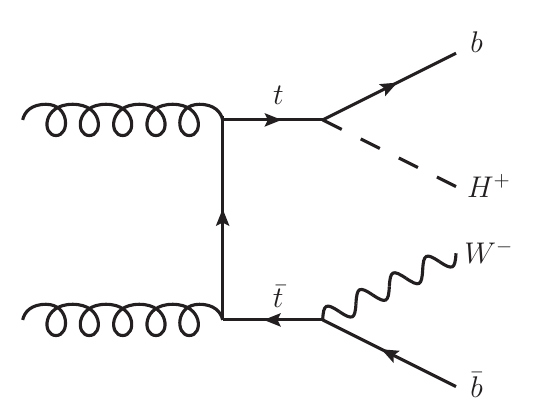}
    \caption{double-resonant top-quark production}
    \label{fig:placeholder1}
  \end{subfigure}
  \caption{Examples of leading-order Feynman diagrams contributing to the production of charged Higgs bosons in pp collisions: (a) single-resonant top-quark production that dominates at large $H^+$ masses, (b) double-resonant top-quark production that dominates at low $H^+$ masses. }
  \label{fig:two-panels2}
\end{figure}

In the feasible parameter space, which satisfies both the theoretical and experimental limits mentioned above, we select six BPs given in Table~\ref{t:BPs}. As seen in the table, the charged Higgs boson is light with mass between $120$ and $300$ GeV. Our study focuses on charged Higgs scalars lighter than the top quark, as required by process $t \rightarrow H^+ b$.
In this study, our signal is given by $pp \rightarrow t\bar{t} \rightarrow H^+ W^- b\bar{b} \rightarrow \ell^- \bar{\nu}b \bar{b}\bar{b}c$ as in Fig.~\ref{fig:placeholder1} and the dominant SM background processes are $t\bar{t}$, $tW^-$ and $t\bar{t}b\bar{b}$, where $j$ corresponds to a light quark or gluon~\cite{ATLAS:2023bzb,Kao:2011aa}. We use the TOP++ package~\cite{Czakon:2011xx} to correct the production cross section of top quark pairs.

For $H^+$ masses below the top-quark mass ($m_{H^+} < m_{\text{top}}$), the main production mechanism is through the decay of a top-quark, $t \to b H^+$. For $H^+$ masses above the top-quark mass ($m_{H^+} > m_{\text{top}}$), the leading production mode is
$gg \to tbH^+$ (single-resonant top-quark production) as in Fig.~\ref{fig:placeholder} and the dominant decay is $H^+ \to tb$, the total decay chain is $pp \to \bar{t}bH^+ \to W^- b\bar{b}tb$. Furthermore, this approach holds for masses lighter than the top quark mass~\cite{ATLAS:2018gfm}. As shown in Fig.~\ref{fig:placech1}, the cross section of single-resonant top-quark production process is very small and lies well below current experimental limits. Therefore, we will not discuss this process in detail but instead focus on the double-resonant top-quark production process, which has a significantly larger cross section.  

\subsubsection{Event generation and selection}

Before starting to discuss our analysis method, we present the toolbox used to generate both signal and background events. At the parton-level, these events are generated using the code \texttt{MadGraph5\_aMC@NLO-3.6.2} \cite{Alwall:2014hca} and employing the NN23LO1 parton distribution functions from the
Lhapdf repository~\cite{Buckley:2014ana}, where we adopt the two following kinematic cuts (in pseudorapidity $\eta$, transverse momentum $p_T$, cone separation and  Missing Transverse Energy $E_T^{\text{miss}}$):
\begin{align}
|\eta(\ell,{j})| < 2.5, \
p_T({j}) > 20~\text{GeV}, \
p_T(\ell) > 10~\text{GeV}, \
\Delta R(\ell\ell/{jj}) > 0.4, \
E_T^{\text{miss}} > 10.
\label{pc1}
\end{align} 
\vspace*{-1.1truecm}

The generating parton-level events are then passed to \texttt{Pythia-8.3}~\cite{Sjostrand:2014zea,Bierlich:2022pfr} to include parton showering, hadronisation and hadron decays. For detector simulation, we utilise \texttt{Delphes-3.5.0}~\cite{deFavereau:2013fsa} with the standard ATLAS card, where we adopt the \texttt{anti-kt} jet algorithm~\cite{Cacciari:2008gp} with a jet parameter $\Delta R = 0.4$. We calculated the final effective cross sections using MadAnalysis 5~\cite{Conte:2012fm}.

We first list in Table~\ref{t:parton_ch} the cross sections of the signal and background processes after applying the parton-level cuts illustrated in Eq.~(\ref{pc1}).
  \begin{table}[!ht]
    \centering
    \scriptsize
    \begin{tabular}{|c|c|c|c|c|c|c|c|c|c|c|c|} 
        \hline
        $\sigma$ (fb) & BP1 & BP2 & BP3 & BP4 & BP5 & BP6 & $t\bar{t}$ & $tW^-$ & $t\bar{t}b\bar{b}$ \\
        \hline
        Parton level & 1525 & 410.2 & 719.8 & 1230 & 2855 & 5934 & 72090 & 160 & 4482 \\
        \hline
Effective cross section & 109.14 & 29.36 & 51.51 & 86.59 & 204.32 & 174.67 & 3486.27 & 0.056 & 405.53 \\
        \hline
    \end{tabular}
    \caption{The signal and background for all signal BPs at the parton level as well as effective cross sections for charged Higgs.}
    \label{t:parton_ch}
\end{table}


To reduce the backgrounds events, we apply the b-tagging and similarly define a c-tagging to be used in subsequent reconstruction. For this reason, we separate signal and background events into three categories: 3b1c (three $b$-jets, one c-jets), 3b1j (three $b$-jets, one light jets), 4b (four $b$-jets) and 2b1c1j (two $b$-jets, one c-jets, one light jets).

\begin{table}[htbp]
\centering
\begin{scriptsize}
\begin{tabular}{|l|c|c|c|c|c|c|c|c|c|c|}
\hline
 & BP1 & BP2 & BP3 & BP4 & BP5 & BP6 & $t\bar{t}$ & $tW^-$  & $t\bar{t}b\bar{b}$ \\
\hline
3b1c & 30.26 & 6.09 & 13.44 & 26.72 & 62.42 & 57.431 & 439.71 & 0.006  & 135.79 \\
\hline
3b1j & 20.03 & 4.13 & 9.80 & 20.09 & 45.38 & 44.58 & 389.11 & 0.004 & 112.08 \\
\hline
4b & 2.75 & 0.49 & 1.94 & 3.18 & 10.47 & 10.45 & 10.47 & 0  & 30.37 \\
\hline
2b1c1j & 57.04 & 16.33 & 27.05 & 49.35 & 109.18 & 103.50 & 2988.98 & 0.044 & 293.82 \\
\hline
\end{tabular}
\end{scriptsize}
\caption{The signal and backgrounds effective cross sections after the pre-selection cuts of charged Higgs.}\label{t:cross_section_bp23}
\end{table}


In Table~\ref{t:cross_section_bp23}, the effective cross sections for both signal and background processes across the aforementioned categories are summarized in this analysis. Overall, all observed cross sections remain relatively small. However, except for the 4b region, the other signal regions maintain comparatively large cross sections. The 4b region, in contrast, presents significant detection challenges. This suppression is primarily due to the strong dependence of lepton reconstruction efficiency and b-tagging performance on the transverse momenta of the reconstructed objects. Additionally, the presence of undetectable particles, such as neutrinos, complicates the evaluation of $E_T^{\text{miss}}$, thereby further hindering accurate event reconstruction.

\subsection{Pseudoscalar Higgs Collider phenomenology}\label{sec4-2}

Neutral pseudoscalar bosons A arise in various extensions of the SM, exhibiting distinct couplings to up- and down-type fermions as well as to gauge bosons. At the LHC, the ATLAS and CMS collaborations have performed searches in b-quark, dimuon, and $\tau\tau$ final states. Among these, the $\tau\tau$ channel is particularly important due to its high lepton identification purity and more reliable background estimation compared to b-quark final states, while also offering higher branching fractions than the dimuon channel because of the larger $\tau$ lepton mass~\cite{ATLAS:2016ivh,CMS:2022goy}. Accordingly, we focus on the process $pp\rightarrow A \rightarrow \tau^+ \tau^-$. 

\subsubsection{Event generation and selection} 
The following acceptance selection cuts are adopted for the signal and background generations:
\begin{align}
|\eta(\ell,{j})| < 2.5, \
p_T({j}) > 20~\text{GeV}, \
p_T(\ell) > 10~\text{GeV}, \
\Delta R(\ell\ell/{jj}) > 0.4, \
E_T^{\text{miss}} > 10.
\label{pc2}
\end{align} 

A veto is applied to remove events with extra electrons or muons that meet looser selection criteria than the corresponding $\tau\tau$ channel definitions. This ensures the mutual exclusivity of the selected event samples. These requirements also help with the suppression of background processes, such as $Z/\gamma^\ast \rightarrow ee$ or $Z/\gamma^\ast \rightarrow \mu\mu$.

We have computed the cross sections for both the signal and background processes at the parton level and effective cross section after applying detector acceptance and selection cuts. The background actually includes both Process $pp\rightarrow \tau^+ \tau^-$ and Process $pp\rightarrow  \tau^+ \tau^- j$, during which the CKKW method~\cite{Alwall:2014hca} was employed for merging.

\begin{table}[htbp]
	\centering
		\begin{scriptsize}
			\begin{tabular}{|c|c|c|c|c|c|c|c|c|c|c|} 
\hline
$\sigma$ (fb) & BP1 & BP2 & BP3 & BP4 & BP5 & BP6 &$\tau^+ \tau^-$  \\ 
\hline
Parton level & 8.238 & 5.295 & 4.528 & 1.157 & 20.06 & 8.238 &1084000\\ 
\hline
Effective cross section & 0.458 & 0.294 & 0.252 & 0.064 & 1.115 & 0.468 &791.32\\ 
\hline
\end{tabular}
		\end{scriptsize}
		\caption{The signal and background for all signal BPs at the parton level as well as effective cross sections of the pseudoscalar boson.}\label{t:parton_ah}
\end{table} 


\subsection{$Z^\prime$ Collider phenomenology}\label{sec4-3}

Searches for additional neutral gauge bosons can be performed in a variety of processes. If such bosons couple to the SM quarks, they may be directly observed through their production and subsequent decay into high energy lepton pairs or jets. The case of the decay into leptons is particularly attractive due to low backgrounds and good momentum resolution~\cite{CMS:2021ctt}. Bounds on several models containing extra neutral gauge bosons have been set by both the CMS experiments by measuring high energy lepton pair production cross sections~\cite{CMS:2022uul}. Searches have been made in the $pp\rightarrow Z^\prime \rightarrow e^+ e^-$ channel, which has the best acceptance. However, since no new particles have been discovered so far, experiments still impose stringent upper limits upon the cross section for $Z^\prime$ production~\cite{CMS:2021ctt}.

\subsubsection{Event generation and selection}
We adopt the following kinematic cuts to improve the efficiency of the MC event generation:
\begin{align}
|\eta(\ell,{j})| < 2.5, \
p_T({j}) > 20~\text{GeV}, \
p_T(\ell) > 10~\text{GeV}, \
\Delta R(\ell\ell/{jj}) > 0.4, \
E_T^{\text{miss}} > 10.
\label{pc3}
\end{align} 

We have computed the cross sections for both the signal and background processes at the parton level and effective cross section after applying detector acceptance and selection cuts. The background actually includes both Process $pp\rightarrow e^+ e^-$ and Process $pp\rightarrow e^+ e^- j$, during which the CKKW method was employed for merging.

  \begin{table}[htbp]
	\centering
		\begin{scriptsize}
			\begin{tabular}{|c|c|c|c|c|c|c|c|c|c|c|} 
\hline
$\sigma$ (fb) & BP1 & BP2 & BP3 & BP4 & BP5 & BP6 &$e^+ e^-$ \\ 
\hline
Parton level & 0.2083 & 0.8152 & 0.2963 & 0.0535& 0.1361 & 0.00460 & 283.1 \\ 
\hline
Effective cross section & 0.0042 & 0.016 & 0.0034 & 0.0011& 0.016 & 0.0005 & 4.73 \\ 
\hline
\end{tabular}
		\end{scriptsize}
		\caption{The signal and background for all signal BPs at the parton level as well as effective cross sections of $Z^\prime$.}\label{t:parton_z}
\end{table}

\section{Signal-to-background analysis}\label{sec5}

\subsection{Event reconstruction}\label{sec5-1}
To enhance signal-background discrimination, we reconstruct key kinematic features of signal events. For the process $pp \rightarrow t\bar{t} \rightarrow H^+ W^- b\bar{b} \rightarrow \ell^- \bar{\nu}b \bar{b}\bar{b}c$,  
which involves a multi-particle decay chain, reconstruction faces significant challenges. The top quark decays to a  b-quark and a charged Higgs boson, while the anti-top decays to a b-quark and a W boson. Each W boson subsequently decays leptonically, producing a charged lepton and an undetectable neutrino. Reconstructing the neutrino momentum via $E_T^{\text{miss}}$ is essential for event interpretation.

\begin{figure}[htbp]
  \centering
  \begin{subfigure}[t]{.30\textwidth}
    \centering
    \includegraphics[width=\linewidth]{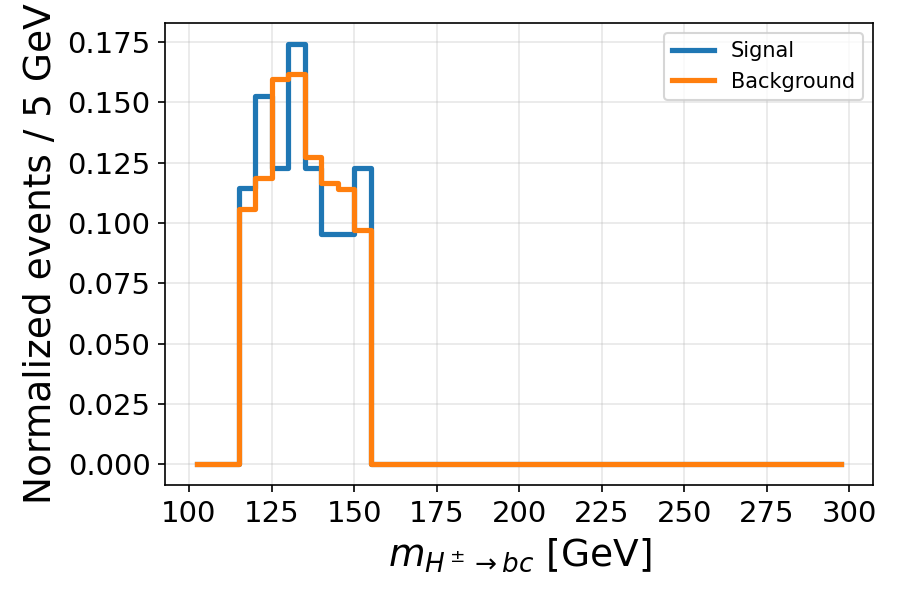}
    \caption{}
    \label{fig:pla1}
  \end{subfigure}
\begin{subfigure}[t]{.30\textwidth}
    \centering
    \includegraphics[width=\linewidth]{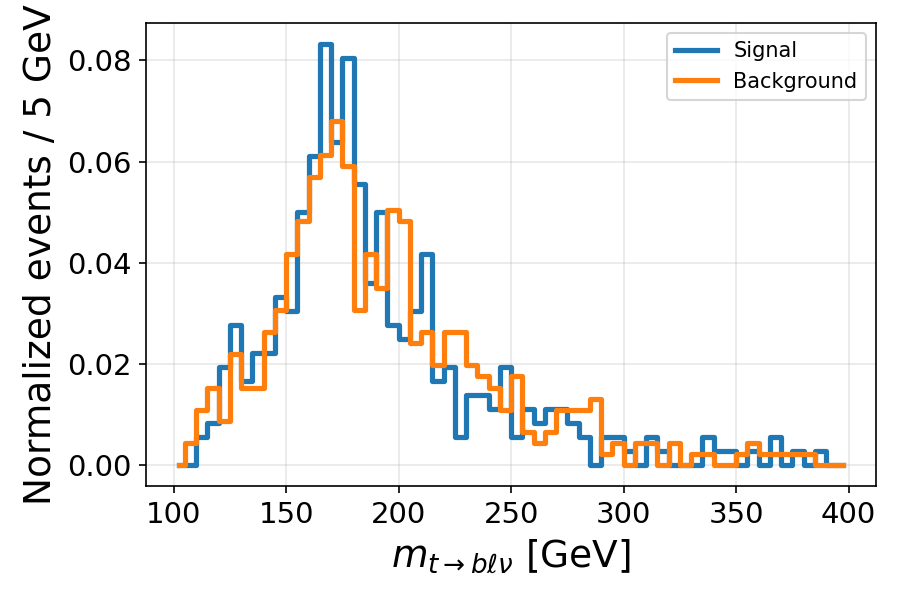}
    \caption{}
    \label{fig:pla2}
 \end{subfigure}
 \begin{subfigure}[t]{.30\textwidth}
    \centering
    \includegraphics[width=\linewidth]{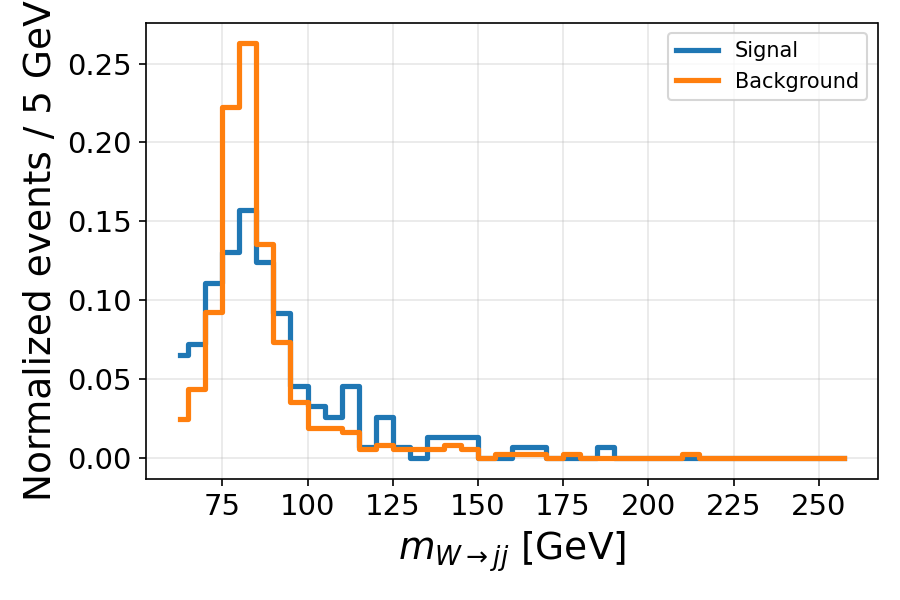}
    \caption{}
    \label{fig:pla3}
 \end{subfigure}
 \begin{subfigure}[t]{.30\textwidth}
    \centering
    \includegraphics[width=\linewidth]{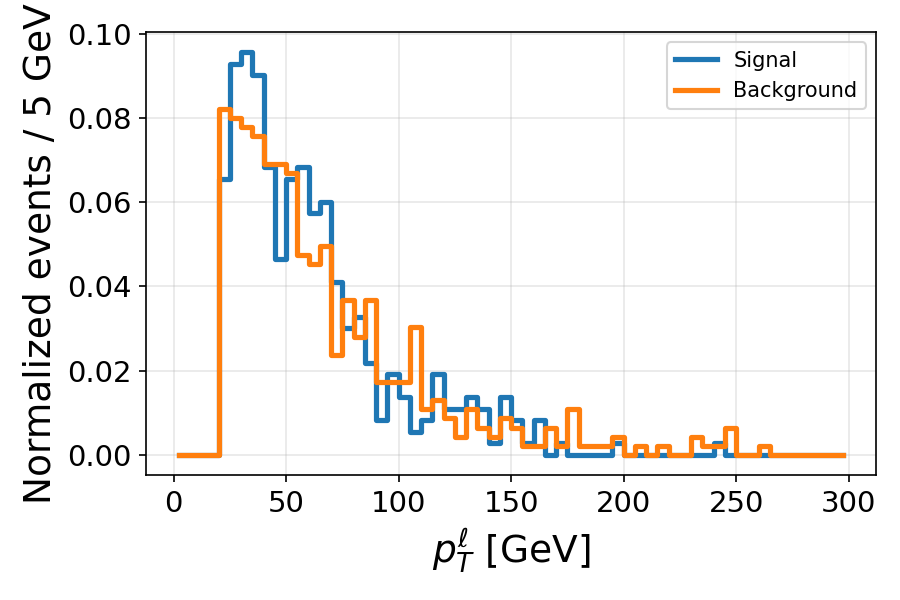}
    \caption{}
    \label{fig:pla4}
 \end{subfigure}
 \begin{subfigure}[t]{.30\textwidth}
    \centering
    \includegraphics[width=\linewidth]{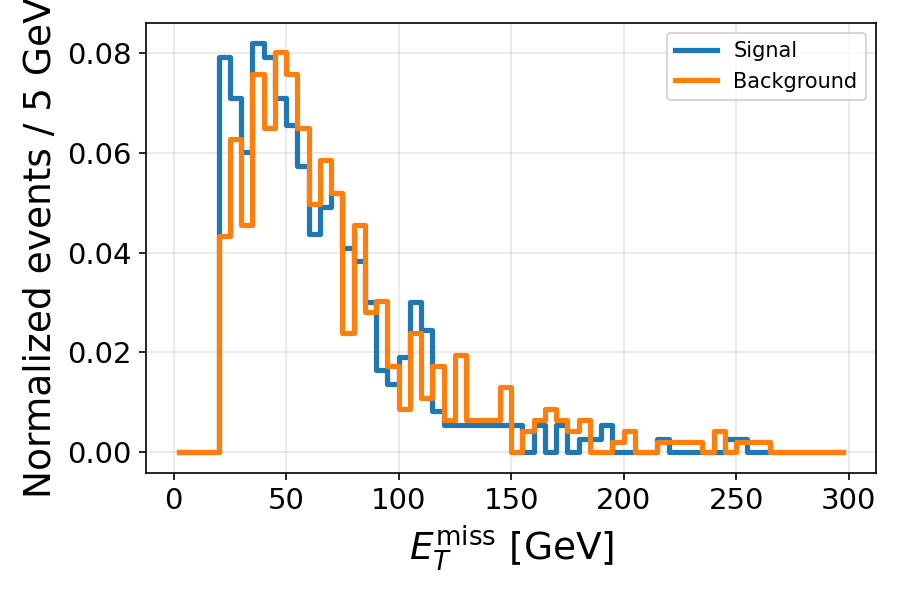}
    \caption{}
    \label{fig:pla5}
 \end{subfigure}

    \caption{The reconstructed (a) charged Higgs boson mass, (b) top quark mass, (c) W boson mass, (d) Lepton transverse momentum, (e) $E_T^{\text{miss}}$. The
signal events are generated by using BP5 in Table ~\ref{t:BPs}.}
  \label{fig:five-panels}
\end{figure}

To address these challenges, we apply a hypothesis-driven reconstruction approach that utilizes the event's $E_T^{\text{miss}}$ to infer momentum from undetected neutrinos. Despite generally small $E_T^{\text{miss}}$ values, we identify the optimal light-jet pair for W boson reconstruction by minimizing the $\chi^2$ deviation between their invariant mass and the known W mass. The resulting reconstructed W boson mass distribution for both signal and background peaks consistently near the true W boson mass.

\begin{figure}[!htbp]
  \centering
  \begin{subfigure}[t]{.40\textwidth}
    \centering
    \includegraphics[width=\linewidth]{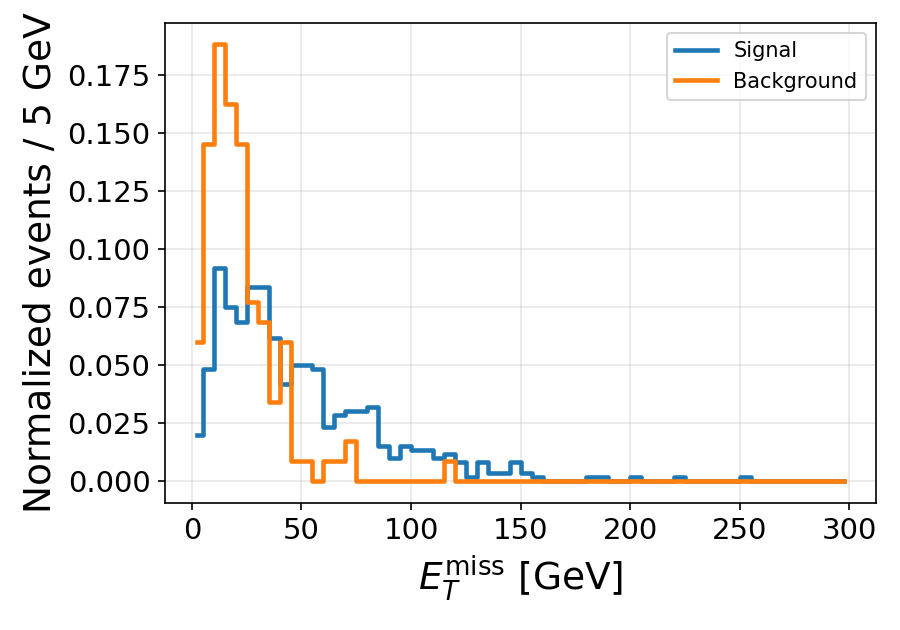}
    \caption{}
    \label{fig:placeh1}
  \end{subfigure}
   \hspace{0.04\textwidth}
\begin{subfigure}[t]{.40\textwidth}
    \centering
    \includegraphics[width=\linewidth]{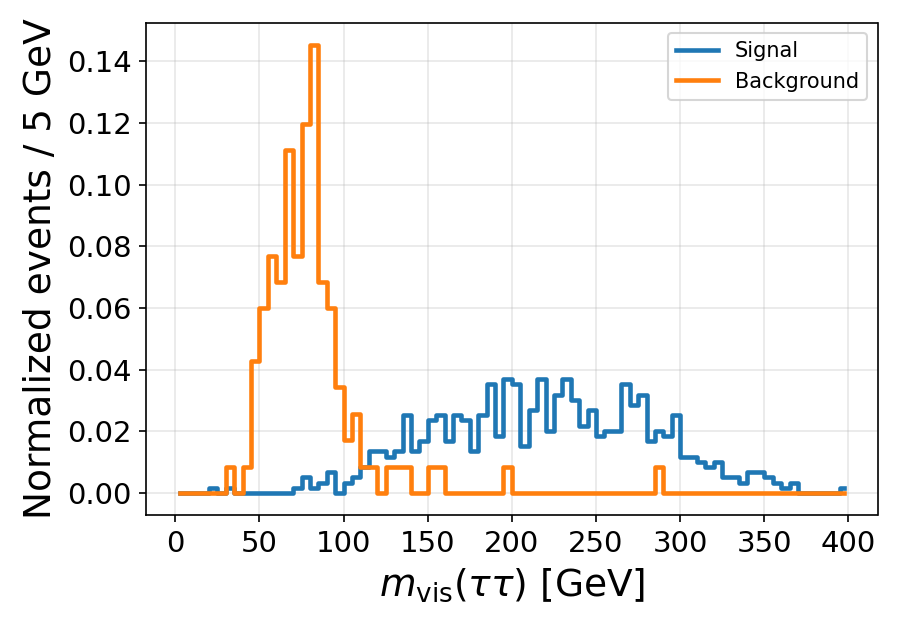}
    \caption{}
    \label{fig:placeh2}
 \end{subfigure}
 \begin{subfigure}[t]{.40\textwidth}
    \centering
    \includegraphics[width=\linewidth]{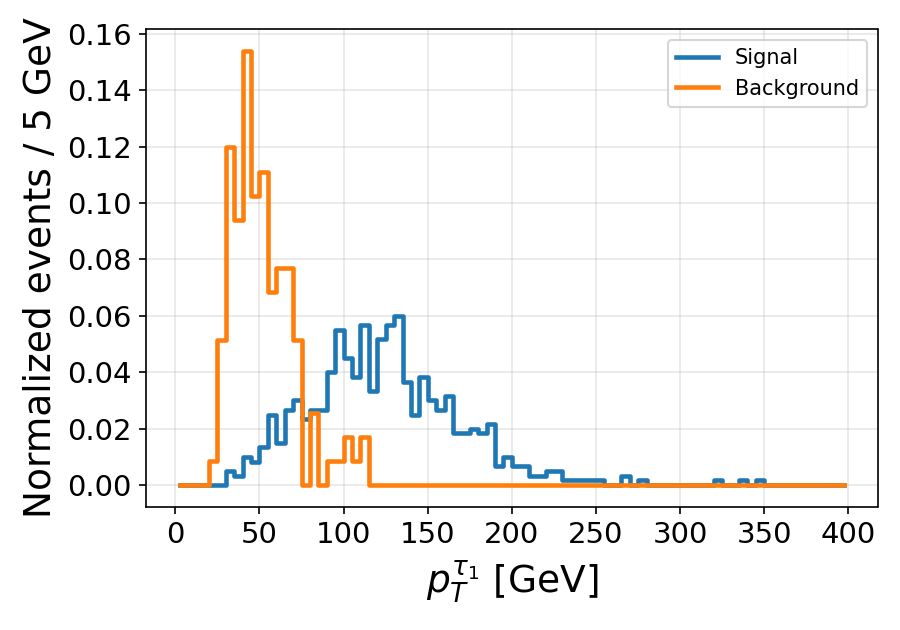}
    \caption{}
    \label{fig:placeh3}
 \end{subfigure}
 \hspace{0.04\textwidth}
 \begin{subfigure}[t]{.40\textwidth}
    \centering
    \includegraphics[width=\linewidth]{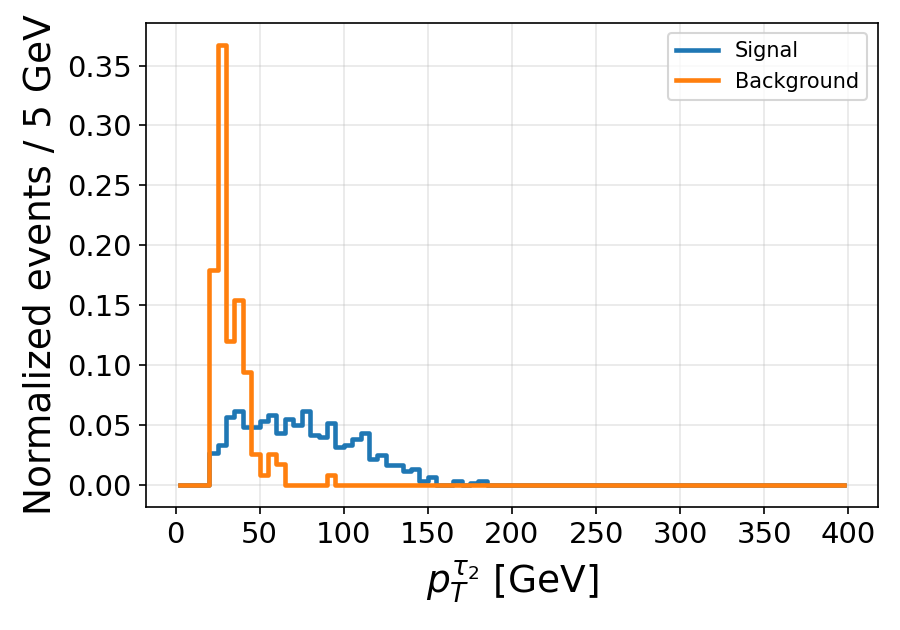}
    \caption{}
    \label{fig:placeh4}
 \end{subfigure}
     \caption{The reconstructed (a) $E_T^{\text{miss}}$, (b) pseudoscalar Higgs boson mass, (c) $\tau_1$ transverse momentum, (d) $\tau_2$ transverse momentum. The signal events are generated by using BP5 in Table ~\ref{t:BPs}.}
  \label{fig:two-pahs}
\end{figure}

The top quark mass is reconstructed from the lepton, b-jet, and neutrino momentum derived from $E_T^{\text{miss}}$. A simplified neutrino approximation $p_x = p_x = E_{Tx}^{\text{miss}}$, $p_y = 0$, $p_z = 0$ is used due to its indeterminate direction. The optimal b-jet is selected via $\chi^2$ minimization, while $H^\pm$ reconstruction combines one b-jet and one c-jet with invariant mass closest to its target value. Energy-momentum conservation constraints ensure physical consistency throughout the reconstruction.

\begin{equation}
\chi^2=(m^{jj}-m_W)^2+(m_{bc}-m_H^\pm)^2+ (m_{b\ell\nu}-m_t)^2
\end{equation}

For the pseudoscalar Higgs boson, the total momentum of the system is calculated by applying opposite-charge selection criteria and summing the momenta of the two $\tau$ particles, under the assumption that the $\tau$ particles are treated as undecayed. A more accurate mass estimation is reconstructed by assuming the particles move collinearly, which is used to estimate the relationship between the two $\tau$ particles and the $E_T^{\text{miss}}$. The visible mass $m_{\text{vis}}$  of the system is then derived from the two $\tau$ particles.

\begin{figure}[!htbp]
  \centering
  \begin{subfigure}[t]{.40\textwidth}
    \centering
    \includegraphics[width=\linewidth]{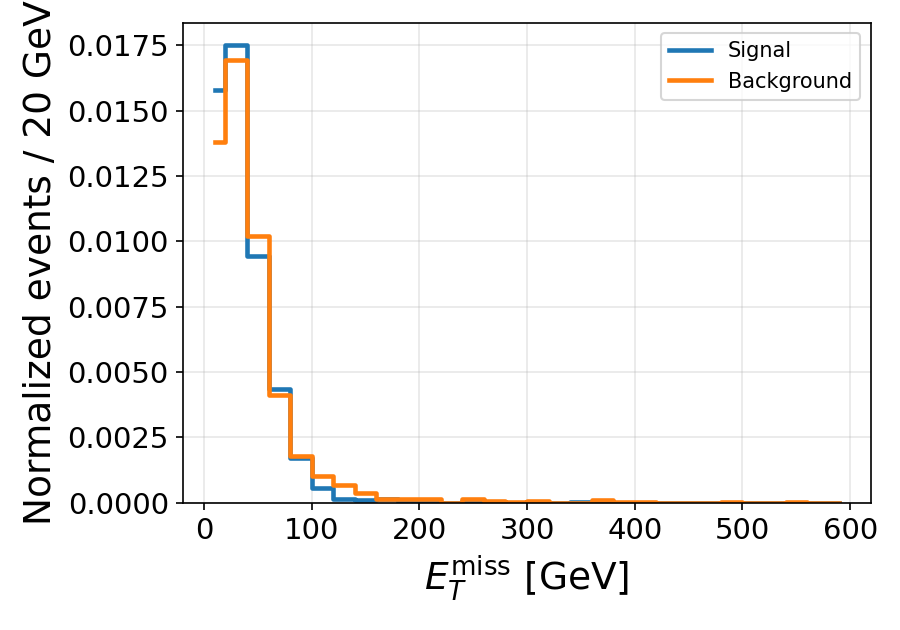}
    \caption{}
    \label{fig:place1}
  \end{subfigure}
   \hspace{0.04\textwidth}
\begin{subfigure}[t]{.40\textwidth}
    \centering
    \includegraphics[width=\linewidth]{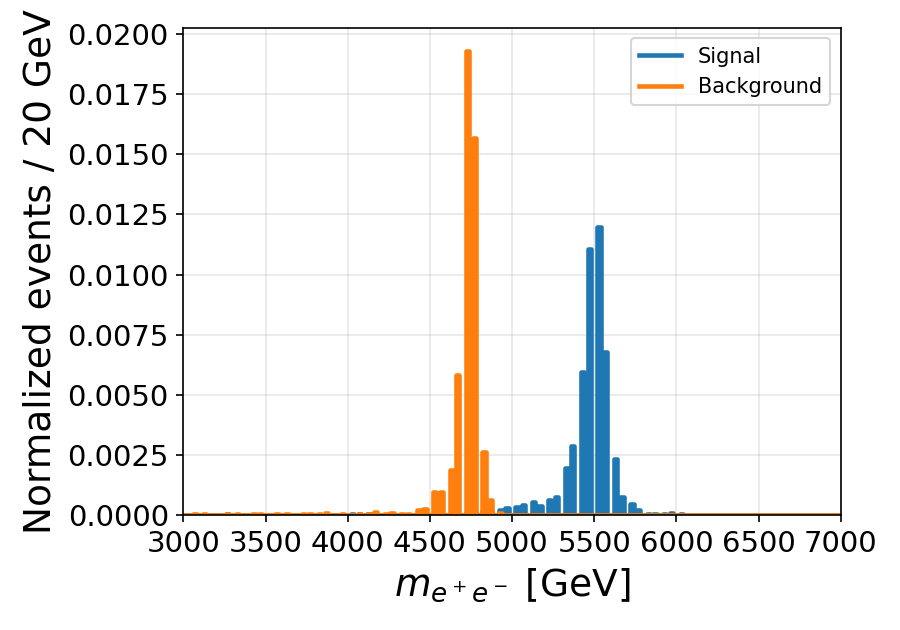}
    \caption{}
    \label{fig:place2}
 \end{subfigure}
 \begin{subfigure}[t]{.40\textwidth}
    \centering
    \includegraphics[width=\linewidth]{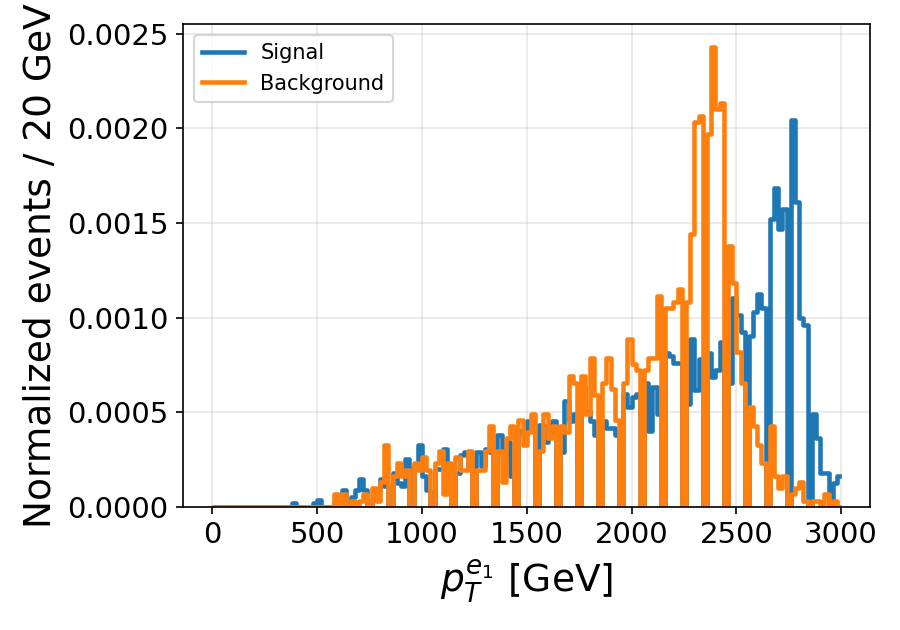}
    \caption{}
    \label{fig:place3}
 \end{subfigure}
 \hspace{0.04\textwidth}
 \begin{subfigure}[t]{.40\textwidth}
    \centering
    \includegraphics[width=\linewidth]{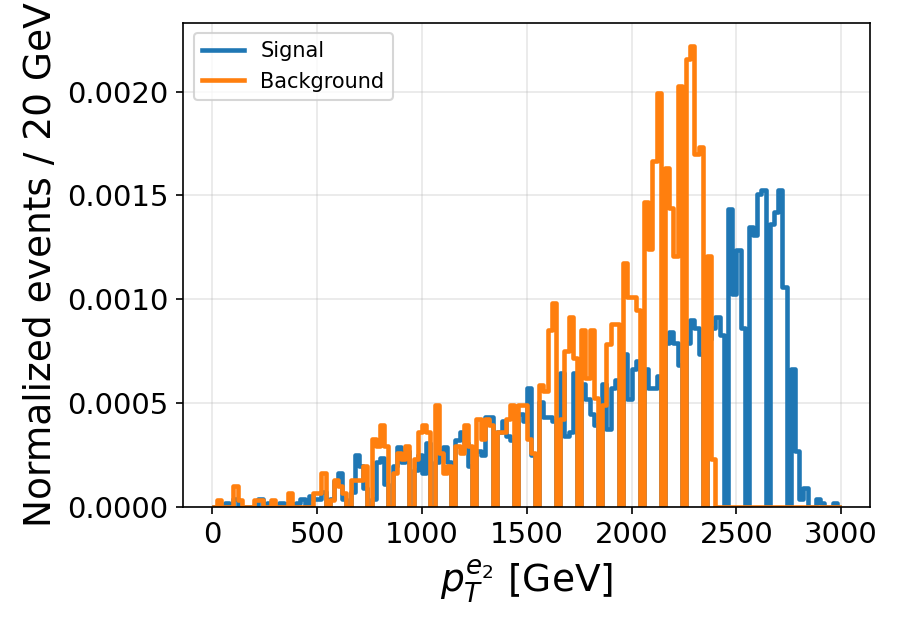}
    \caption{}
    \label{fig:place4}
 \end{subfigure}
     \caption{The reconstructed (a) $E_T^{\text{miss}}$, (b) $Z^\prime$ boson mass, (c) $e_1$ transverse momentum, (d) $e_2$ transverse momentum. The
signal events are generated by using BP5 in Table ~\ref{t:BPs}.}
  \label{fig:two-pas}
\end{figure}

As shown in Fig.~\ref{fig:placeh2}, the observed differences between the signal and background primarily arise from the fact that the intermediate particles in the background are predominantly $Z/\gamma^\ast$. The discrepancy in the transverse momenta of the two $\tau$ particles reflects the constraints imposed by momentum and energy conservation.

The reconstruction method for the $Z^\prime$ boson is very similar to that of the pseudoscalar Higgs boson, as we employed nearly identical strategies. The $E_T^{\text{miss}}$ is small, and the peak in the invariant mass of the background primarily originates from other heavy scalar particles. The transverse momenta of the two electrons are significantly large due to the high mass of the intermediate particle.

\subsection{Significances}\label{sec5-2}

We are using the MadAnalysis 5 recast module by PAD database~\cite{Conte:2014zja} and Spey package~\cite{Araz:2023bwx} to investigate the relationship between theoretical calculations and experimental data in order to determine the exclusion of theories by experiments. For the $Z^\prime$ boson, use the ATLAS-EXOT-2018-30 card; for the pseudoscalar Higgs boson, use the ATLAS-EXOT-2018-30 card. However, for the specific charged Higgs boson process under investigation, no suitable parameter card is currently available in PAD database. 

\begin{figure}
    \centering
    \includegraphics[width=0.5\linewidth]{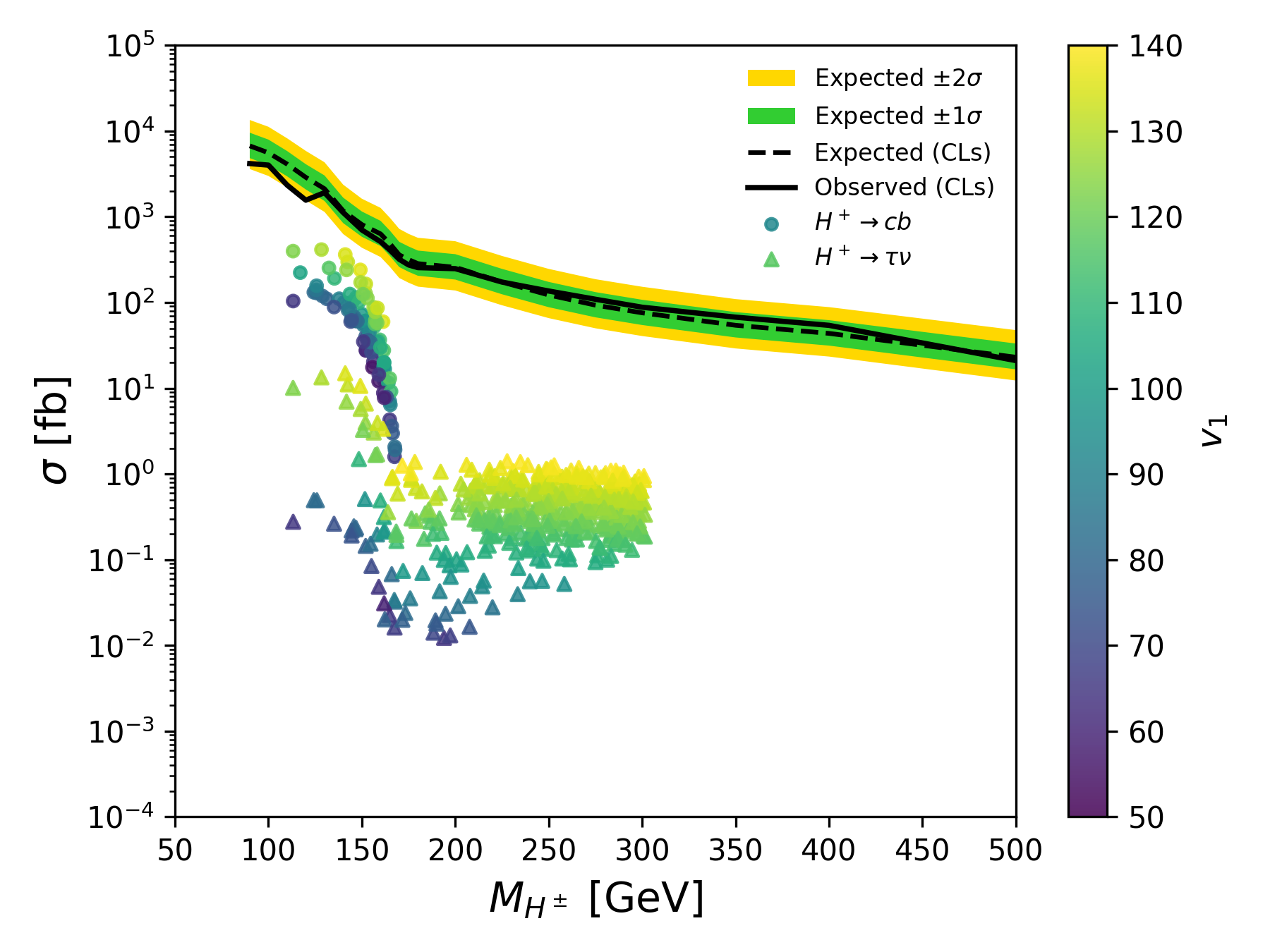}
    \caption{Observed and expected 95\% CL exclusion limit on cross section as a function of the charged Higgs boson mass in 36.1 $\mathrm{fb}^{-1}$ of pp collision data at $\sqrt{s} = 13 \text{TeV}$. Dots represent the cross sections of Process $pp \rightarrow t\bar{t} \rightarrow H^+ W^- b\bar{b} \rightarrow \ell^- \bar{\nu}b \bar{b}\bar{b}c$, while triangles represent the cross sections of Process $pp \to \bar{t}bH^+ \to W^- b\bar{b}tb$.}
    \label{fig:placech1}
\end{figure}

Fig.~\ref{fig:placech1} shows the cross section of the lightest charged Higgs boson as functions of $v_1$ and mass, with exclusion limits from Ref.~\cite{ATLAS:2018gfm} applied. The circles represent processes $pp \rightarrow t\bar{t} \rightarrow H^+ W^- b\bar{b} \rightarrow \ell^- \bar{\nu}b \bar{b}\bar{b}c$, while the triangles correspond to the $pp \to \bar{t}bH^+ \to W^- b\bar{b}tb$ channel. Although both processes remain below current experimental bounds, the cross section for $pp \rightarrow t\bar{t} \rightarrow H^+ W^- b\bar{b} \rightarrow \ell^- \bar{\nu}b \bar{b}\bar{b}c$ is significantly larger, particularly at lower charged Higgs masses and larger $v_1$, thus justifying our selection of benchmark points below the top quark mass to enhance detection prospects at the LHC.

\begin{figure}[htbp]
  \centering
  \begin{subfigure}[t]{.40\textwidth}
    \centering
    \includegraphics[width=\linewidth]{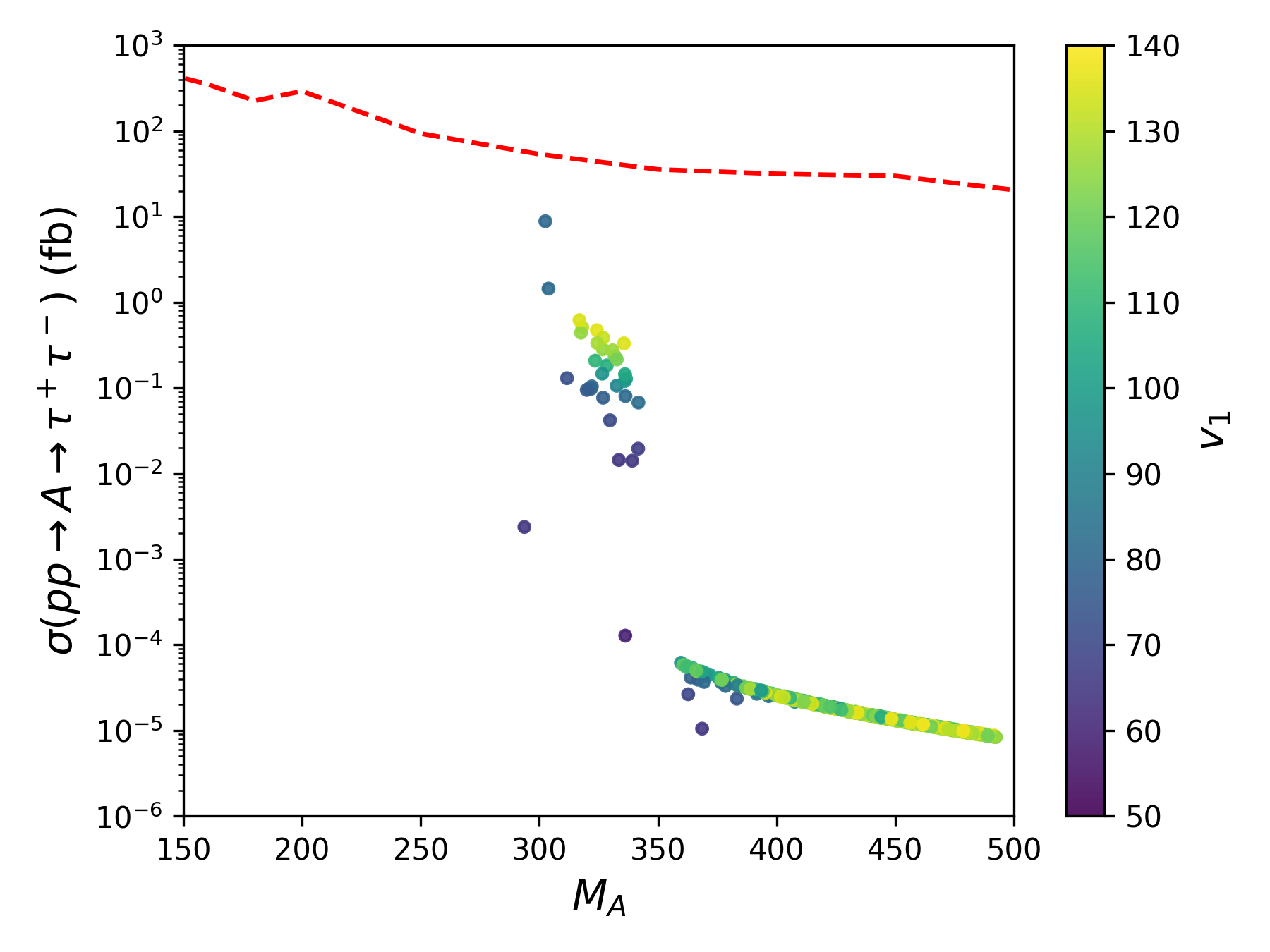}
    \caption{}
    \label{fig:placeah1}
  \end{subfigure}
   \hspace{0.04\textwidth}
\begin{subfigure}[t]{.40\textwidth}
    \centering
    \includegraphics[width=\linewidth]{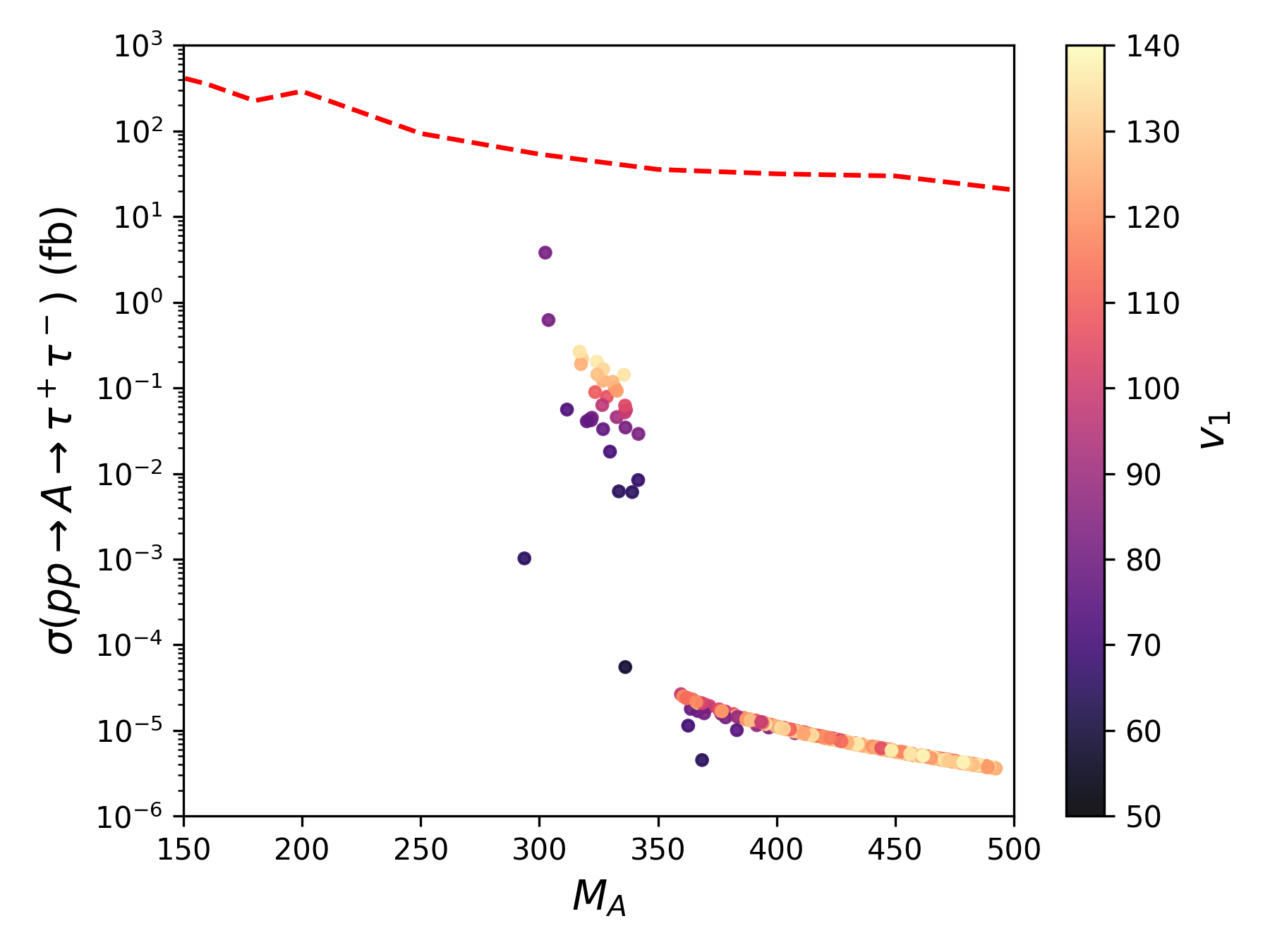}
    \caption{}
    \label{fig:placeah2}
 \end{subfigure}    

    \caption{(a) Cross section as a function of the pseudoscalar Higgs particle massusing a data sample collected with the CMS detector at $\sqrt{s} = 13 \text{TeV}$, corresponding to an integrated luminosity of 138 $\mathrm{fb}^{-1}$. (b) Cross section as a function of the pseudoscalar Higgs particle by recast condition.}
  \label{fig:two-panels}
\end{figure}

In Fig.~\ref{fig:placeah1}, we show the cross section of the lightest pseudoscalar Higgs versus $v_1$ and mass. The red dashed line indicates the experimental exclusion limit~\cite{CMS:2022goy}, and circles represent cross sections from our parameter scan. For masses below 350 GeV, cross sections are relatively large, occasionally exceeding experimental bounds. Above 350 GeV, they decrease nearly linearly with mass. At fixed mass, larger $v_1$ yields higher cross sections due to reduced $v_3$, which enhances the relevant Yukawa couplings and vertex interactions. In Fig.~\ref{fig:placeah2}, we present the recast cross-section for $pp\rightarrow A \rightarrow \tau^+ \tau^-$ using the parameter ATLAS-EXOT-2018-30 card, which exhibits a smaller value, indicating the need for more precise experiments to probe pseudoscalar Higgs within the FDM.

\begin{figure}[htbp]
  \centering
  \begin{subfigure}[t]{.40\textwidth}
    \centering
    \includegraphics[width=\linewidth]{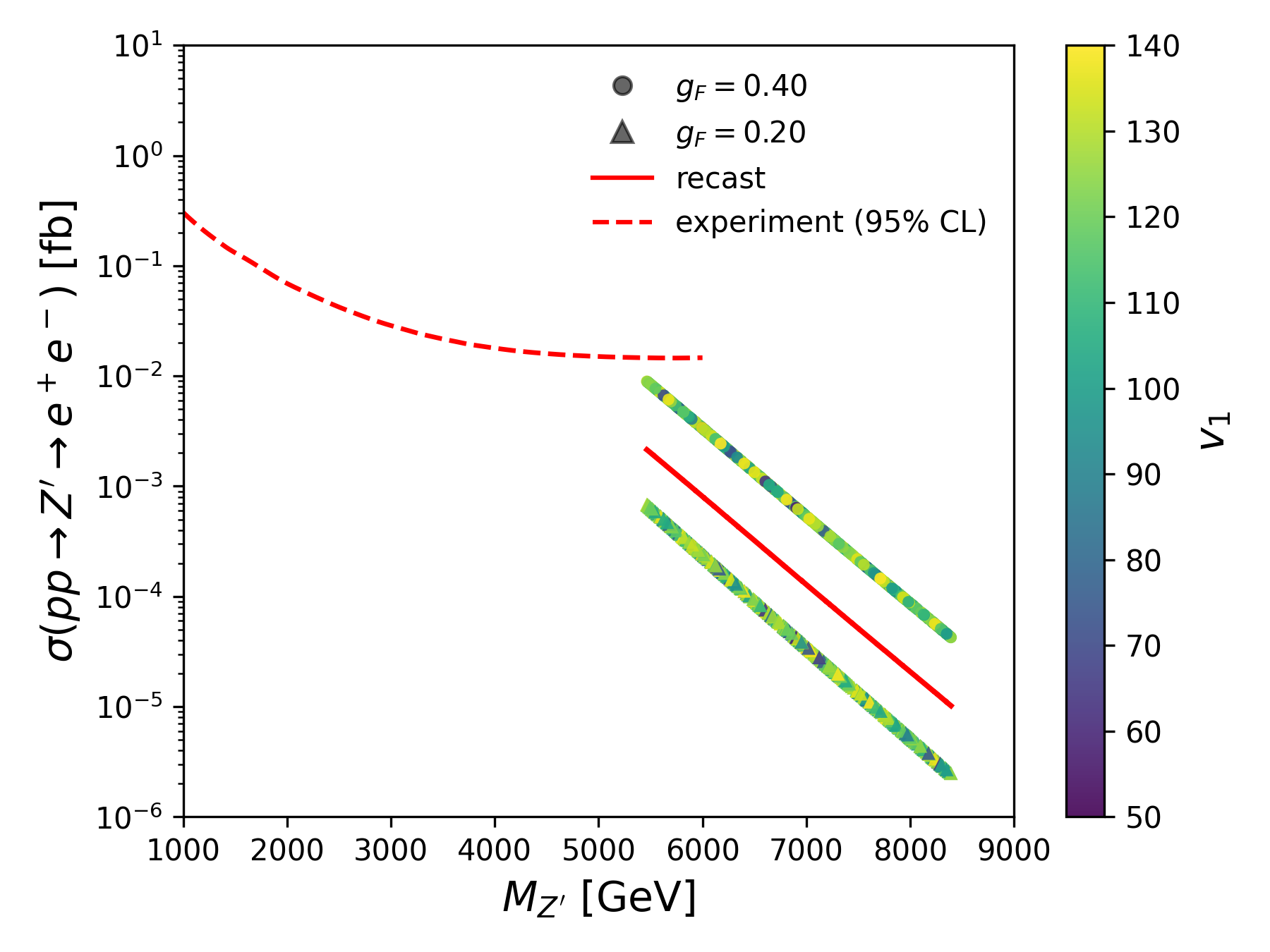}
    \caption{}
    \label{fig:placezp}
  \end{subfigure}
   \hspace{0.04\textwidth}
  \begin{subfigure}[t]{.40\textwidth}
    \centering
    \includegraphics[width=\linewidth]{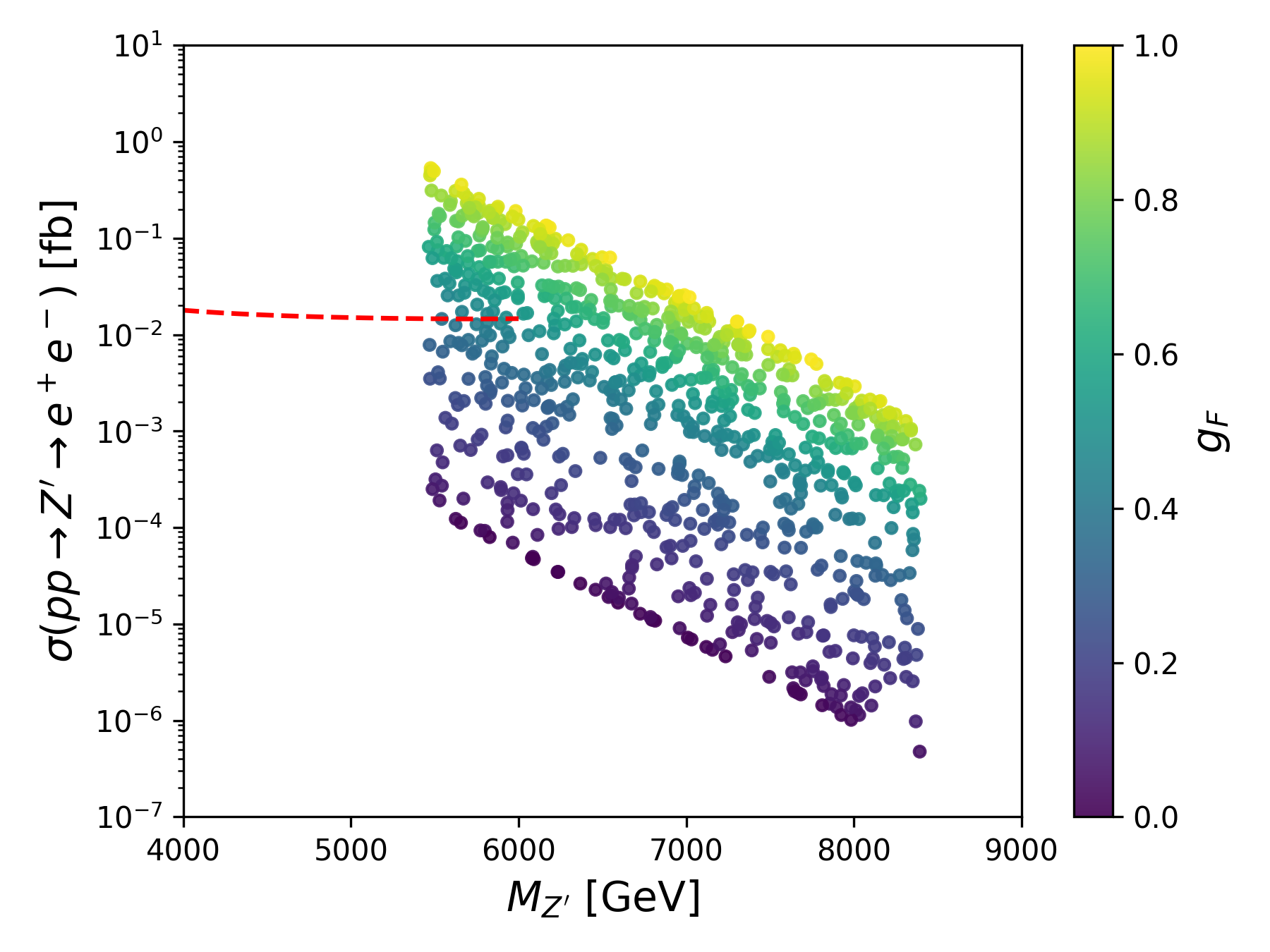}
    \caption{}
    \label{fig:placezp1}
  \end{subfigure}\hfill   

    \caption{Cross section of $pp\rightarrow Z^\prime \rightarrow e^+ e^-$ as a function of the $Z^\prime$ mass measured in proton-proton collisions by the CMS experiment at the LHC. Data were collected at $\sqrt{s} = 13\ \mathrm{TeV}$ from 2016 to 2018, corresponding to an integrated luminosity of up to $140\ \mathrm{fb}^{-1}$. (a) Dots represent results for $g_F = 0.4$, while triangles correspond to $g_F = 0.2$, and the red solid line indicates the recast cross-section. (b) The cross section scaling linearly with $g_F$.}
  \label{fig:two-panels1}
\end{figure}

In Fig.~\ref{fig:placezp}, we show the cross section of $Z^\prime$ boson versus $v_1$ and $Z^\prime$ mass. The red dashed line indicating the experimental exclusion limit~\cite{CMS:2021ctt}. Circles and triangles denote $g_F = 0.40$ and $0.20$, respectively. The red solid line indicates the recast cross-section obtained with the parameter card from ATLAS-EXOT-2016-32 at $g_F = 0.40$.  The $Z^\prime$ cross section decreases nearly linearly with its mass. Fig.~\ref{fig:placezp1} shows the cross section scaling linearly with $g_F$ (scanned from 0 to 1). When $g_F < 0.40$, all predicted values lie below current experimental limits over the studied mass range. If the recast constraints are applied, the limit can be relaxed to $g_F < 0.60$.

 To better investigate the detection prospects for these particles at the future 300 $\mathrm{fb}^{-1}$ collider, we examined the corresponding discovery significance and exclusion significance using six benchmark points as examples. We use the PYHF package~\cite{Feickert:2022lzh} and use Minuit optimizer to perform a likelihood analysis by fitting the observed data distribution to a signal-plus-background hypothesis, among the observation data uses the asimov-sb distribution sampling method. This approach eliminates the need for tedious MC toy simulations. 

\begin{table}[htbp]
	\begin{center}
	\centering
		\begin{scriptsize}
			\begin{tabular}{|c|c|c|c|c|c|c|c|c|c|c|} 
\hline
$H^\pm$ & BP1 & BP2 & BP3 & BP4 & BP5 & BP6 \\ 
\hline
$Z_{obs}$ & 1.56 & 0 & 0.298 & 1.21 & 3.29 & 2.79  \\ 
\hline
$Z_{exp}$ & 1.65 & 0.447 & 0.823 & 1.38 & 3.21 & 2.75  \\ 
\hline
		\end{tabular}
		\end{scriptsize}
		\caption{Discovery significance of $H^\pm$ are generated by using signal and background in Table ~\ref{t:parton_ch}.}\label{t:parton_chh}
	\end{center}
\end{table} 
This set of six benchmark points for the charged Higgs boson $H^\pm$ overall does not reach the "discovery" threshold. The expected sensitivity $Z_{exp}$ increases progressively from BP1 to BP6, indicating enhanced analysis sensitivity with varying parameter masses. The observed significance $Z_{obs}$ largely follows this trend: BP5 shows a local excess of approximately 3.29$\sigma$, almost consistent with the expected 3.21$\sigma$, while BP6 also lies around 2.79$\sigma$, reflecting normal statistical fluctuations within the highest sensitivity region. In contrast, BP2 and BP3 are significantly lower than expected, resembling downward background fluctuations. Overall, the current data do not support a significant new physics signal. At most, BP5 can be regarded as a local ~3$\sigma$ hint, though its significance would generally decrease when considering the look-elsewhere effect and global significance.

We further validate the performance under the conditions of the HL-LHC operating at an integrated luminosity of 3000 $\mathrm{fb}^{-1}$.
\begin{table}[htbp]
	\begin{center}
	\centering
		\begin{scriptsize}
			\begin{tabular}{|c|c|c|c|c|c|c|c|c|c|c|} 
\hline
$H^\pm$ & BP1 & BP2 & BP3 & BP4 & BP5 & BP6 \\ 
\hline
$Z_{obs}$ & 5.17 & 0 & 0.941 & 3.81 & 10.04 & 8.80  \\ 
\hline
$Z_{exp}$ & 5.46 & 1.48 & 2.59 & 4.34 & 10.01 & 8.67  \\ 
\hline
		\end{tabular}
		\end{scriptsize}
		\caption{Discovery significance of $H^\pm$ are generated by using signal and background in Table ~\ref{t:parton_ch} in HL-LHC.}\label{t:parton_chlh}
	\end{center}
\end{table} 

As shown in Fig.~\ref{t:parton_chlh}, the HL-LHC at an integrated luminosity of 3000 $\mathrm{fb}^{-1}$ holds strong potential for discovering indications of a charged Higgs boson. The findings point toward a mass most likely below 145 GeV, with the data at 135 GeV showing a high discovery confidence. This provides supporting evidence for the existence of a charged Higgs boson in the vicinity of 130 GeV.

\begin{table}[htbp]
	\begin{center}
	\centering
		\begin{scriptsize}
			\begin{tabular}{|c|c|c|c|c|c|c|c|c|c|c|} 
\hline
$A$ & BP1 & BP2 & BP3 & BP4 & BP5 & BP6 \\ 
\hline
$Z_{obs}$ & 0 & 0 & 0.0015 & 0 & 0.0021 & 0.001  \\ 
\hline
$Z_{exp}$ & 0.0166 & 0.0107 & 0.0092 & 0.0023 & 0.0405 & 0.017  \\ 
\hline
		\end{tabular}
		\end{scriptsize}
		\caption{Discovery significance of $A$ are generated by using signal and background in Table ~\ref{t:parton_ah}.}\label{t:parton_ahh}
	\end{center}
\end{table} 
 Overall, this set of six benchmark points for the pseudoscalar Higgs boson shows neither evidence of discovery nor meaningful sensitivity: all $Z_{obs}$ values are approximately zero, and the expected significances $Z_{exp}$ are also extremely small (the maximum is only 0.0405$\sigma$ for BP5). This indicates that the signal is difficult to detect under such conditions, primarily due to the overwhelming background. Achieving observable significance would require designing specialized cuts criteria for the collider analysis.

\begin{table}[htbp]
	\begin{center}
	\centering
		\begin{scriptsize}
			\begin{tabular}{|c|c|c|c|c|c|c|c|c|c|c|} 
\hline
$Z^\prime$ & BP1 & BP2 & BP3 & BP4 & BP5 & BP6 \\ 
\hline
$Z_{obs}$ & 0.0184 & 0.091 & 0.0185 & 0 & 0.091 & 0  \\ 
\hline
$Z_{exp}$ & 0.078 & 0.030 & 0.006 & 0 & 0.03 & $0$  \\ 
\hline
\end{tabular}
		\end{scriptsize}
		\caption{Discovery significance of $Z^\prime$ are generated by using signal and background in Table ~\ref{t:parton_z}.}\label{t:parton_zh}
	\end{center}
\end{table}

Overall, this set of six $Z^\prime$ benchmark points exhibits negligible significance: both observed and $Z_{exp}$ are at the level of 0 to 0.1$\sigma$ (the slight excesses of ~0.09$\sigma$ in BP2 and BP5 are only minor upward fluctuations), which is fully consistent with pure background. This also indicates that the current analysis has almost no sensitivity to these benchmark points. To make progress, further reduction of background contributions is necessary rather than solely increasing luminosity or energy.

\section{Conclusions}\label{sec6}

In this paper, we investigate the production and decay processes of the lightest charged Higgs boson $H^\pm$, the lightest pseudoscalar Higgs boson A, and the $Z^\prime$ boson in the FDM model at the LHC detector. Specifically, for the lightest charged Higgs boson, we examine two distinct detection processes based on single-resonant top-quark production or double-resonant top-quark production. We employ mainstream Monte Carlo simulation tools to simulate the production and decay of these particles. Reconstruction and recasting analyses demonstrate that existing experimental conditions are insufficient to conclusively detect these particles. To further explore the detection prospects at future colliders with a center-of-mass energy of 14 $\text{TeV}$ and an integrated luminosity of 300 $\mathrm{fb}^{-1}$, we analyze six benchmark points for signal and background processes. The results indicate that the $Z^\prime$ and pseudoscalar Higgs boson at these masses have very low discovery significance. This suggests that detection remains challenging and requires the design of specific search strategies and optimized selection cuts, or perform a comprehensive scan and mass spectrum fitting. The results for the charged Higgs boson support the experimentally observed ~3$\sigma$ excess around 130 \text{GeV}, which is highly likely to be probed at the HL-LHC with an integrated luminosity of 3000 $\mathrm{fb}^{-1}$.

\section*{Acknowledgments}

The work has been supported by the National Natural Science Foundation of China (NNSFC) with Grants No. 12075074, No. 12235008, Hebei Natural Science Foundation for Distinguished Young Scholars with Grant No. A2022201017, No. A2023201041, Natural Science Foundation of Guangxi Autonomous Region with Grant No. 2022GXNSFDA035068, and the youth top-notch talent support program of the Hebei Province.

\end{document}